\documentclass[prb,groupedaddress,showpacs,twocolumn]{revtex4}
\usepackage{graphicx}

\def\figwidth{\columnwidth}
\begin{document}
\title{Pinned Wigner Crystals}
\author{R. Chitra}
\email{chitra@lptl.jussieu.fr}
\affiliation{LPTL, Universite de Pierre et Marie Curie, Jussieu, Paris-75005, France }
\author{T. Giamarchi}
\email{giam@lps.u-psud.fr} \affiliation{Laboratoire de Physique
des Solides, CNRS-UMR 85002, UPS Bat. 510, 91405 Orsay France}
\author{P. Le Doussal}
\email{ledou@lpt.ens.fr} \affiliation{CNRS-Laboratoire de
Physique Theorique de l'Ecole Normale Superieure, 24 rue Lhomond,
75231 Cedex 05, Paris, France.}
\date{\today}
\begin{abstract}
We study the effects of weak disorder on a Wigner crystal in a magnetic field.
We show that an elastic description of the pinned Wigner crystal provides
an excellent framework to obtain most of the physically relevant observables.
Using such a description, we compute the static and dynamical properties.
We find that, akin to the Bragg glass phase,
a good degree of translational order survives (up to a large lengthscale in $d=2$, infinite in $d=3$).
Using a gaussian variational method, we obtain the full frequency dependence of the conductivity tensor.
The zero temperature Hall resistivity
is independent of frequency and remains unaffected by disorder at its classical value.
We show that the characteristic features of the conductivity in the pinned Wigner crystal
are dramatically different from those arising from the naive extrapolations of
Fukuyama-Lee type theories for charge density waves.
We determine the relevant scales and find that
the physical properties depend crucially on whether the disorder correlation length is
larger than the cyclotron length or not.
We analyse, in particular, the magnetic field and density dependence of the optical conductivity.
Within our approach the pinning frequency can increase with increasing magnetic field
and varies as $n^{-3/2}$ with the density $n$.
We compare our predictions with recent experiments on transport in two dimensional
electron gases under strong magnetic fields. Our theory allows for a consistent interpretation of
these experiments in terms of a pinned WC.
\end{abstract}
%\maketitle must follow title, authors, abstract and \pacs
\maketitle
%\tableofcontents

\section{Introduction}
Amongst the various effects of strong interactions in electronic systems,
the possibility of crystallization of electrons, initially predicted by Wigner
\cite{wigner_crystal}, is one of the most exciting.
For densities lower than a critical density $n_c$,
the Coulomb potential energy dominates over the kinetic
energy and one expects the formation of a Wigner crystal (WC), where the
electrons occupy the sites of a lattice (triangular in two dimensions).
A measure of this critical density is
the dimensionless parameter $r_s$ defined as the ratio of the Coulomb
to Fermi energies. Unfortunately, crystallization requires
extremely low densities which are hard to obtain for three
dimensional systems. The situation is much better for a two dimensional
electron gas where
Monte-Carlo simulations  \cite{ceperley_qmc_wigner}
have shown that the formation of a WC requires a density corresponding to
 $r_s \ge 37$. First, it is easier to access  such densities and  second,
the value of $n_c$ can be increased by the application of a
strong magnetic field  applied perpendicular to the plane of the 2DEG.
Two dimensional electron gases under strong magnetic fields  are thus the
prime candidate  systems in which to observe the phenomenon of
 Wigner crystallization.

Indeed an insulating state was observed in experiments on monolayer systems
\cite{andrei_wigner_2d,willett_wigner_resistivity,goldman_wigner_threshold,williams_wigner_threshold}.
This insulating state was characterized by a diverging diagonal resistivity $\rho_{xx}$  for  temperature
$T\to 0$ and activated behavior for finite $T$. Moreover,
the Hall resistivity $\rho_{xy}$  was found to be temperature independent and nearly equal to its classical value.
For the pure systems, approximate calculations \cite{cote_energy_2deg_short,cote_energy_2deg_long} have
shown that the WC becomes the lowest energy state when the  filling factor $\nu$ which is
the ratio of the density to the magnetic field equals $\nu \le 1/5$ for  Ga-As electron systems
and around $\nu= 1/3$ for the  hole  like systems. So it is quite reasonable to interpret
the insulating state as a Wigner crystal pinned by the
incipient disorder in the sample.  This is consistent
with  the diverging linear resistivity and  with measurements of
nonlinear $I-V$ characteristics. Nonetheless, given the absence of direct imaging of the system,
the unambiguous identification
of this phase as a WC was the subject of intense debate. Also, the theoretical analysis
was complicated by various important factors compared to the simple case proposed by Wigner:
(i) in these systems in strong magnetic fields,  the phase diagram is very rich, since the interplay
of interaction and disorder can lead to phases such as the integer and fractional
quantum Hall effect. The   non-crystalline phase is not a free electron gas but a strongly
correlated phase. It is therefore, not obvious that one does not need a description anchored in the
quantum Hall physics to describe the insulating phase as well \cite{zhang_hall_insulator};
(ii) disorder modifies drastically the physical properties of crystalline phases.
This makes
the determination of the transport properties which is our only probe of the WC crystal so far,
much more difficult to compute.

On the theoretical side, the question of how to  describe such pinned crystalline phases
is still open. Our understanding of the effect of disorder on such elastic structures
stemmed from the pioneering works of Larkin \cite{larkin_70,larkin_ovchinnikov_pinning} for vortices
and Fukuyama and Lee (FL)\cite{fukuyama_cdw_pinning} for charge density waves,
which predict that the crystal gets pinned by
disorder and perfect translational order is lost. As a result of the pinning,
 the  a.c. transport
develops a peak at a disorder dependent pinning frequency $\omega_p$. However, due to the  qualitative nature of these theories, very little was known beyond these qualitative aspects of a pinned system.
This is a tremendous handicap
for the  WC since, contrarily to CDW, the problem of a WC involves many length scales.
Moreover, the loss of  translational order  and the possibility of defects led to doubts about
the validity  of   an elastic theory  to study the transport.
Recent experiments\cite{li_conductivity_wigner_magneticfield,hennigan_optical_wigner,beya_thesis,li_conductivity_wigner_density} have
started to obtain the full frequency dependent conductivity, and the magnetic field and density
dependence of observables such as the pinning frequency. The fact that these
experiments did not comply
with simple FL type expectations, cast doubts on the interpretation of this experimental phase as a
pinned WC. This clamors for  a quantitative theory of pinned WC that goes beyond the FL approximations,
and with which experiments can be compared to.

Fortunately, the recent  progress made in the understanding of such disordered elastic structures
\cite{giamarchi_vortex_long,giamarchi_columnar_variat,giamarchi_book_young,giamarchi_quantum_revue}
presents us with methods, which go beyond
simple scaling level arguments, to tackle periodic systems with disorder. These techniques were used to show
\cite{giamarchi_vortex_long} that
despite pinning, a classical three dimensional periodic system is defect free and
retains quasi long range translational order (Bragg glass).
In this paper, we use these methods to obtain a {\it quantitative} theory of the pinned WC.
We show that an elastic description is an excellent approximation.
We compute both the static and the a.c. transport properties and investigate in
detail the magnetic field dependence of the various relevant quantities, such as
the pinning frequency and threshold field. We show that
the behavior of a pinned WC has marked differences from  the one that was commonly
accepted, based on the analogies with pinned charge density waves. Our theory
permits  a reasonably consistent description of the existing experimental results, and
suggests additional ways to check for a pinned WC through the transport properties.
A summary of the method and some of the results on the a.c. transport were presented in a shorter form in
Ref.~\onlinecite{chitra_wigner_hall}.

The plan of the paper is as follows.
In Sec.~\ref{sec:modelsys}, we define the elastic Hamiltonian  used to study this
system, and discuss the relevant length-scales.
The Gaussian variational method (GVM) that we use to solve
the problem, and  the variational equations are described in Sec.~\ref{sec:gvm}. In Sec.~\ref{sec:solution},
we obtain the solution of these equations. These two sections contain most
of the technical parts of the paper, and  we  concentrate on the
physical results in the remaining sections. The reader interested only in the
physical consequences of our theory can skip these two sections and jump
directly to Sec.\ref{sec:static}. In this section, we discuss the static properties
of the pinned WC, i.e. the compressibility and the translational order.
The dynamics is examined in Sec.~\ref{sec:dynamics}. General features of the
conductivity are described, together with the magnetic field and density dependence
of the a.c. transport. Our theory is compared with existing experiments on the
pinned WC in Sec.~\ref{sec:experiments}, while we discuss other theories in
Sec.~\ref{sec:otherth}. Conclusions can be found in Sec.~\ref{sec:conclusion}
 and many of the
 technical details have been relegated to the Appendices.

\section{Modeling  the system} \label{sec:modelsys}

\subsection{Elastic model} \label{sec:elasticmodel}

\begin{figure}
\centerline{\includegraphics[width=\figwidth]{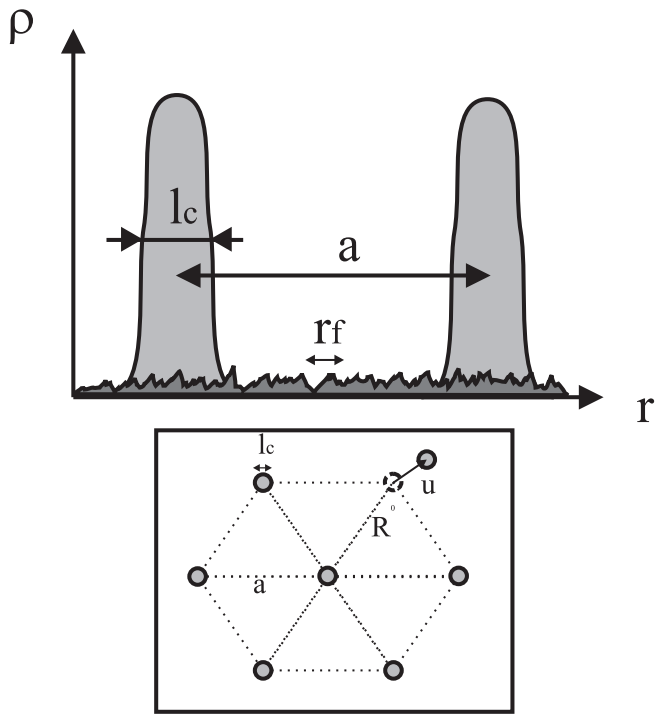}}
\caption{\label{fig:basiclength} The three length characterizing the
Wigner crystal. The size $l_c$ of the ``particles'' in the crystal (at low temperature
it is essentially given by the extension
of the wavefunction around the equilibrium position and is the cyclotron radius, at large
temperatures it is controlled by the thermal fluctuations and is the Lindemann length), $a$ the lattice spacing is
controlled by the density of particles, and the disorder is correlated over
a length $r_f$. The inset shows  the triangular structure of the Wigner crystal. Particles
are labeled by an equilibrium position ${ R}_i$ and a displacement ${ u}$.}
\end{figure}

Let us define the model we use to study the insulating crystal phase.
Firstly, we see that
retaining  the full fermionic nature of the wavefunction, would lead
to a very complicated  problem in the presence of disorder and
coulomb interactions. Fortunately,
in the crystal phase the particles become localized, and thus
distinguishable. So a ``classical'' description for the crystal is a
good starting point. The quantum aspects of the original fermionic
problem will be hidden in the parameters characterizing this crystal.
In the crystalline phase, the electrons occupy the sites of a triangular lattice with a lattice
constant $a$ which is related to the density of electrons by $n \sim (\pi a^2)^{{-1}}$ .
As shown in Figure~\ref{fig:basiclength}, the electrons at site $i$ are displaced from
their mean  equilibrium positions labeled by the two dimensional vector
${  R}_i$, by $u(R_i,t)$. The ``effective'' size of the particles in this crystal are
determined by the spatial extension of the electronic wavefunction, as shown
in Figure~\ref{fig:basiclength}. In the presence of a  strong magnetic field, this
is simply the cyclotron radius $l_c=\sqrt{1/(eB)}$.  At finite temperatures, thermal
fluctuations add to this purely quantum extension of the wavefunction. Hence the
``size'' of the particle gets renormalized. For high temperatures, it tends to the
Lindemann length $l_T$, defined by the average thermal displacement
$l_T^2 = \langle u^2 \rangle$. A rough estimate of the size of the particle is
thus $\sqrt{l_c^2 + l_T^2}$. In the following, we will simply denote this
(temperature dependent) length by $l_c$.
Note that this kind of  classical
approach is certainly valid for strong magnetic fields where, $l_c \ll a$.
In this case, the overlap of the various electronic wavefunctions is
negligible and the Fermi statistics can  be  safely ignored. It is rather difficult to give a precise
estimate of the field and density for which this approach breaks down.
A useful hint is nonetheless given by the Lindemann criterion for the melting of a classical crystal
$l_c^2 \sim C_L^2 a^2$, where the Lindemann number $C_L \sim 0.1$. This  shows that even
displacements close to melting remain relatively small. It is thus reasonable to expect
that our approach would remain valid even relatively close to melting.
Our approach also applies to Wigner crystal at $B=0$, provided  the correct
extension of the wavefunction is used in place of $l_c$.

The long wavelength
properties of the WC can now be described by an elastic
hamiltonian describing the displacements $u$. We set $\hbar=1$ in the rest of the paper. The vibration modes of this
crystal can be quantized in the usual manner leading to the imaginary time
elastic action in the presence of disorder,
\begin{eqnarray}\label{eq:ham}
S &=& \frac12 \int_{\bf  q} \frac1\beta \sum_{\omega_n}
u_{\alpha}(q,\omega_n)[(\rho_m\omega_n^2 +c q^2)\delta_{\alpha
\beta} + d \frac{q_\alpha q_\beta}{q}  \nonumber \\
& &  + i \rho_m\omega_n \omega_c  \epsilon_{\alpha\beta}] u_{\beta}(-q,-\omega_n) \\
& & +  \int d^{2}r \int_0^{\beta } d\tau V({ r})\rho({ r},\tau) \nonumber
\end{eqnarray}
where here and below we denote the integration over the Brillouin zone (BZ)
\begin{equation}
\int_{\bf q} = \int_{BZ} \frac{d^2q}{(2\pi)^2}
\end{equation}
and $u_\alpha$ with $\alpha=x,y$ are the two components of the
vector $u$, $\epsilon_{\alpha\beta}$ is the antisymmetric tensor
$\epsilon_{xy}=-\epsilon_{yx}=1$ and the Matsubara frequencies $\omega_n = \frac{2\pi n}{ \beta}$
with the inverse temperature  $\beta=1/T$. $\rho_m \simeq {m \over {\pi a^2}},\rho_c \simeq {e \over {\pi a^2}}$ are
the mass and charge densities. $c$ and $d$ are the shear and
bulk moduli. These moduli can be obtained from an expansion of the
coulomb correlation energy of the WC in terms of the displacements.
Analytical expressions for the elastic moduli were obtained for the
classical crystal in Ref.~\onlinecite{bonsall_elastic_wigner}. Later,  efforts
were made to estimate quantum
corrections to these moduli using trial
wave functions\cite{maki_elastic_wigner}.
Here, we work
with the classical values obtained  in Ref.~\onlinecite{bonsall_elastic_wigner} where
the elastic moduli are independent of the magnetic field, which should be a reasonable
approximation for strong fields. The bulk modulus $d$ and the shear modulus
$c$ are given by
\begin{eqnarray}
d &=& \frac{\rho_c^2}{\epsilon} \\
c &=& \alpha \frac{\rho_c^2a}{\epsilon}
\end{eqnarray}
where  $\alpha \simeq 0.03$ and $\epsilon$ is the dielectric constant of the substrate.
Since the bulk modulus describes compressional modes, it is drastically
affected by the coulomb repulsion. The off-diagonal
term in the action (\ref{eq:ham}) comes from the Lorentz force,
and is proportional to the cyclotron frequency $\omega_c={{\rho_c
B} / {\rho_m}}$, where $B$ is the magnetic field applied
perpendicular to the plane of the crystal. Finally, the last term
describes the coupling  to disorder, modeled here by a random
potential $V$. The density of particles
\begin{equation}
\rho({ r}) = \sum_i \overline{\delta}({ r} - { R}_i - { u}_i)
\end{equation}
where $\overline{\delta}$ is a $\delta$-like function of range $l_c$
(see Figure~\ref{fig:basiclength}) and ${ u}_i \equiv { u}({ R}_i)$.
Since the disorder can vary at a lengthscale $r_f$ {\it a priori} shorter or
comparable to the lattice spacing $a$, the continuum limit ${ u}_i
\to { u}(r)$,
valid in the elastic limit $|{ u}_i - { u}_{i+1}| \ll a$ should be taken with
care in the disorder term \cite{giamarchi_vortex_short,giamarchi_vortex_long}.
This can be done using the
decomposition of the density in terms of its Fourier components
\begin{equation} \label{eq:fourdens}
\rho({ r})\simeq \rho_0 - \rho_0\nabla\cdot { u} +
\rho_0 \sum_{{ K} \neq 0} e^{i { K}\cdot({ r} - { u}(r))}
\end{equation}
where $\rho_0$ is the average
density and ${ K}$ are the reciprocal lattice vectors  of the perfect crystal.
The finite range of $\overline{\delta}$ is recovered\cite{giamarchi_vortex_long}
by restricting the sum over $K$ to momentum of order $K_{\rm max} \sim \pi/l_c$
We assume a gaussian distribution for the disorder $V(r)$, which leads to the
following disorder averages
\begin{equation}
\overline{V({ r}) V({ r}')} = \Delta_{r_f}({ r}-{ r}')
\end{equation}
$\Delta_{r_f}$ is a delta-like function of range $r_f$ which is the characteristic
correlation length of the disorder potential (see Figure~\ref{fig:basiclength}).
The gaussian limit is valid when there are many weak pins. We will come back to
this point in a more quantitative manner in the next section.

\subsection{Physical lengthscales} \label{sec:lengthscales}

The physical properties of the various phases of the two dimensional electron crystal  are completely governed
by the interplay of the the various microscopic  length scales in the system: the lattice spacing $a$,
the particle size $l_{c}$ and the disorder correlation
length $r_{f}$. These microscopic parameters define collective lengthscales
which are  of fundamental importance in the physics of
such disordered systems \cite{giamarchi_vortex_long,giamarchi_book_young,giamarchi_quantum_revue}.
These  length scales, $R_c$ and $R_a$, are the distances over which
the relative displacements are of the order of the size of the particle and
the lattice spacing respectively. They are defined by the disorder averaged displacement
correlation functions
\begin{eqnarray}\label{lengths}
\langle [{ u}({ R}_c) -{ u}({ 0})]^2\rangle
&=& \text{max} [r_f ^2, l_c^2 ] \equiv \xi^2_0 \\ \nonumber
\langle [{ u}({ R}_a) -{ u}({ 0})]^2\rangle
&=& a^2
\end{eqnarray}
In order to distinguish between these two lengths it is mandatory
to keep all the harmonics in the decomposition of the density (\ref{eq:fourdens}).
Keeping only one harmonic, as is done for CDW, amounts to considering that the effective
``size'' of the particle is $a$ and hence $R_c \sim R_a$. For the Wigner crystal,
as we will see, it is crucial to carefully distinguish between these
two lengthscales.

The length $R_a$ can easily be obtained by a static scaling argument comparing the cost
in shear elastic energy and the gain due to disorder.
For the WC, where
$n \simeq (\pi a^{2})^{-1}$ we get
\begin{equation}\label{lengthra}
R_a = {{c a^2} \over {n \sqrt \Delta}}
\end{equation}
In order to be in the weak pinning regime, and for an elastic description to
be valid, it is necessary to have $R_a \gg a$.

The interpretation of $R_c$,
the so called Larkin-Ovchinikov length\cite{larkin_ovchinnikov_pinning} is
more subtle. It  corresponds to
lengths  where the particles can truly feel the random potential. In particular,
it is the length above which the Larkin model \cite{larkin_70}, which approximates the random
potential by a random force becomes invalid, and pinning and metastability appear.
If $R_c \gg a$, the system is collectively pinned, whereas one would have single
particle pinning in the opposite case. We emphasize that the elastic
approximation is valid regardless of the behavior of $R_c$, provided one  still
has $R_a \gg a$.

Since $u(R_a) \sim a$,  $R_a$ has in the past  been incorrectly interpreted  as the lengthscale
at which topological defects such
as dislocations  appear and the translational order of the crystal is lost.
The naive picture is  the one of a crystal broken into crystallites
of size $R_a$. It has recently been shown that this view is grossly incorrect.
In particular, even in two dimension topological defects appear at a lengthscale $R_D$ much greater than $R_a$,
and the crystal preserves a very good degree of positional order \cite{giamarchi_vortex_long,ledoussal_dislocations_2d}.
In three dimensions, the situation is even more favorable
since the system preserves a quasi translational order (power-law divergent Bragg peaks) and
topological defects are not generated by weak disorder ($R_D=\infty$). The
physical implications of these results for the present problem will be discussed in section~\ref{sec:limitations}.

\section{GVM} \label{sec:gvm}

\subsection{Method}

We now study the disordered WC described by (\ref{eq:ham}).
Due to the nonlinear coupling  of the disorder to the displacement field  $u$ in (\ref{eq:ham}),
this problem is extremely difficult to solve. Here, we treat it using a
variational method \cite{giamarchi_columnar_variat,chitra_wigner_hall}.
Many of the technical details
and subtleties of the method can be found in the literature (see e.g.
Ref.~\onlinecite{giamarchi_book_young,giamarchi_quantum_revue} for a review).
We focus directly on the WC and
we present here only the main steps.
The first step involves  averaging  (\ref{eq:ham}) over disorder  by introducing replicas.
This averaging results in an effective action which involves interactions
between the $n$ replicas, given by
\begin{widetext}
\begin{eqnarray}\label{eq:seff}
S &=& \frac12 \sum_{\omega_n} \int_{\bf q}
\sum_a  u_\alpha^a(q,\omega_n)\left[(\rho_m\omega_n^2 +c q^2)\delta_{\alpha \gamma}
+i\rho_m \omega_n\omega_c \epsilon_{\alpha\gamma} + d \frac{q_\alpha q_{\gamma}}{q}
\right] u^a_\gamma(-q,-\omega_n)  \nonumber \\
& & - \frac{\rho_0^2}{2} \int d^2r \int_0^{\beta
}\int_0^{\beta  }d\tau d\tau'
\sum_{a,b,{ K}} \Delta_K \cos\left[K\cdot\left(u^a({ r},\tau) -
u^b({ r},\tau')\right) \right]
\end{eqnarray}
\end{widetext}
where summations over  $\alpha$ and $\beta$  are implicit. The replica
indices, $a,b$  run from $1$ to $n$. ${ K}$ denotes  the  reciprocal lattice vectors.
The size of the particles $l_c$ and the finite
correlation length of the disorder restrict the sum over ${  K}$ to
values of $ K$ smaller than
$K_{\rm max} \sim \pi/\max(r_f,l_c)$. $\Delta_K$ is taken as a constant $\Delta_K = \Delta$
for $K < K_{\rm max}$ and zero otherwise. A more precise formulation is given in
Appendix~\ref{ap:cutdelta}.
The physical disorder averages are recovered in the limit
$n\to 0$.  (\ref{eq:seff}) is quite general and can be used to
describe a host of physical systems, both quantum and classical
\cite{giamarchi_quantum_revue}.

The disorder couples the various replicas via the cosine term. Though
this cosine term is local in space, it is completely non-local in time. This non-locality leads to
a rich
dynamical behavior as we shall show in this paper.
In addition to the cosine term, another quadratic term is generated by
the $q=0$ part of the disorder. This  can be absorbed in a shift of the displacement
field $u$. It affects the static correlation functions but not the transport properties
and thus we neglect it henceforth.

We now search  for a variational solution to  (\ref{eq:seff}) by using
the best quadratic action approximating (\ref{eq:seff}). We use the trial
action
\begin{equation}
S_0= \frac{1}{2 \beta} \int_{{\bf  q}} \sum_n u_{\alpha}^a
(q,\omega_n) G^{{ab}-1}_{\alpha \beta}(q,\omega_n)
u_{\beta}^b(-q, -\omega_n)
\end{equation}
where the whole Green's function $G^{{ab}-1}_{\alpha \beta}(q,\omega_n)$
are variational parameters. The variational free
energy is now given by
\begin{equation}\label{eq:free}
F_{\rm var} = F_0 + \langle S-S_0 \rangle_{S_0}
\end{equation}
The variational parameters are then determined by the saddle point equations
\begin{equation} \label{eq:saddle}
\frac{\partial F_{\rm var}}{\partial G^{{ab}-1}_{\alpha \beta}(q,\omega_n)}=0
\end{equation}
The explicit form of these  equations   are
given in Appendix~\ref{ap:solvar}.

\subsection{Saddle point equations}

The saddle point equations  (\ref{eq:saddle}) have to be solved in the limit of the number of replicas $n\to 0$.
There are two kinds of saddle point solutions in general: one with replica symmetry and
the other with replica symmetry breaking.
For the case of interest here, $d=2$, one can show
that the correct way to take the limit $n\to 0$ is to break the
replica symmetry \cite{giamarchi_columnar_variat}.
We first  introduce the displacement correlation
function
\begin{widetext}
\begin{eqnarray} \label{eq:discor}
B_{\alpha \beta}^{ab}({ x},\tau) &=& \langle [u_\alpha^a ({
x},\tau)-u_\beta^b({ 0},0)]^2 \rangle
=  \frac\beta \int_{\bf  q} \sum_n
\left(G_{\alpha\beta}^{aa}(q,\omega_n)
+G_{\alpha\beta}^{bb}(q,\omega_n) -2\cos(qx +\omega_n
\tau)G_{\alpha\beta}^{ab}(q,\omega_n)\right)
\end{eqnarray}
\end{widetext}
We then parametrize the replica  Green's function as follows:
\begin{equation}\label{eq:param}
[G^{-1}]^{\alpha \beta}_{ab}(q,\omega_n) = f^{\alpha\beta}(q,\omega_n)\delta_{ab} +\sigma^{\alpha \beta}_{ab}(\omega_n)
\end{equation}
where $\sigma$ are the variational parameters and $f^{\alpha\beta}(q,\omega_n)$ is the elastic matrix
defined in (\ref{eq:ham}).
Since the disorder induced interaction
between replicas
is local in space and non-local in time, the parameters $\sigma$ depend only on the frequency.
We  now define  the connected part of the Green's functions as
\begin{equation} \label{eq:connected}
G_c^{\alpha \beta}(q,\omega_n) = \sum_b G^{\alpha \beta}_{ab}(q,\omega_n)
\end{equation}
The saddle point equations, the $f_{\alpha\beta}$ and the various connected Green's functions
are all given in the
 appendix~\ref{ap:solvar}.

The matrix structure of the saddle point equations simplify if we use the
basis of longitudinal and transverse displacements.
This offers a physically
transparent picture in  that the transverse modes are the shear
modes  of the solid and the longitudinal ones represent the
compression modes. In this basis, the Cartesian displacements are given by
\begin{equation}
u_\alpha(q) =  u^L(q)\hat{q}_\alpha  + u^T(q)\epsilon_{\alpha\beta}\hat{q}_\beta
\end{equation}
where $\hat{q} = q/|q|$ is the unit vector along $q$. For the pure system, in the
absence of a magnetic field, the longitudinal and transverse modes are the eigenmodes
of the system. The longitudinal mode is sensitive to the Coulomb repulsion whereas the
transverse mode is not.
The non-disordered part of the Hamiltonian can be rewritten as
\begin{eqnarray}
S[u]& =& \int_ {\bf q}  \sum_n  [ u^L_{q,\omega_n}(\rho_m\omega_n^2 +cq^2
+dq)u^L _{-q,
-\omega_n}\nonumber\\
& & + u^T_{q,\omega_n} (\rho_m \omega_n^2 + c q^2 )u^T_{-q,
-\omega_n} \nonumber \\
 & &+ \rho_m\omega_c \omega_n (u^L_{q,\omega_n} u^T_{-q ,-\omega_n}
-  u^L_{-q ,-\omega_n} u^T_{q ,\omega_n)}]
\end{eqnarray}

The magnetic field mixes these two modes. Note that the longitudinal mode yields  the
correct dispersion for the plasmon mode $\omega \sim q^{1/2}$.
Though the disorder term in (\ref{eq:seff}) has a complicated form in terms of $u_L$ and $u_T$, the
 saddle point equations  themselves can be easily written in
terms of these longitudinal and transverse modes.
  Since, the appropriate solution is the
replica symmetry broken one, all quantities which are off-diagonal in the
replica indices are parametrized in terms of a continuous variable
$0\le u \le 1$. There is a break-point $u_{c}$ such that for $u<u_{c}$ the solutions
manifest explicit replica symmetry breaking and one recovers effectively
replica symmetric solutions for $u>u_{c}$. The saddle point equations, their
parametrisation in terms of
$u$, the structure of the various replica Green's functions are all discussed
elaborately in Appendix~\ref{ap:solvar}.

In this section, we present only the final version of these equations. These equations contain
a parameter $\Sigma$ which can be seen as a disorder induced mass gap and the function
$I(i \omega_{n})$ which describes essentially the dissipation due to disorder.
In terms of these new variables, the various connected Green's functions are
now given by
\begin{widetext}
\begin{eqnarray}
G_{cT}(q,i\omega_n)&=&{{cq^2 + dq+ {\rho_m}\omega_n^2
+I(i\omega_n) +\Sigma(1- \delta_{n,0})} \over {[(cq^2 + dq+
{\rho_m}\omega_n^2 +I(i\omega_n) +\Sigma(1-\delta_{n,0}))(cq^2 +
{\rho_m}\omega_n^2 +I(i\omega_n) +\Sigma(1-\delta_{n,0} ))+
{\rho_m^2}\omega_n^2
\omega_c^2]}} \nonumber\\
G_{cL}(q,i\omega_n)&= &{{cq^2     + {\rho_m}\omega_n^2 +I(i\omega_n)
+\Sigma(1-\delta_{n,0})}\over {[(cq^2 + dq+ {\rho_m}\omega_n^2
+I(i\omega_n) +\Sigma(1-\delta_{n,0}))(cq^2 + {\rho_m}\omega_n^2
+I(i\omega_n) +\Sigma(1-\delta_{n,0} ))+   {\rho_m^2}\omega_n^2
\omega_c^2]}} \label{eq:gfn}\\
G_{cLT}(q,i\omega_n)&= &{{ {\rho_m}\omega_n\omega_c
}\over {[(cq^2 + dq+ {\rho_m}\omega_n^2
+I(i\omega_n) +\Sigma(1-\delta_{n,0}))(cq^2 + {\rho_m}\omega_n^2
+I(i\omega_n) +\Sigma(1-\delta_{n,0} ))+   {\rho_m^2}\omega_n^2
\omega_c^2]}} \nonumber
\end{eqnarray}
\end{widetext}
where $I(i\omega_n)$  satisfies (in the semi-classical limit)
\begin{widetext}
\begin{eqnarray}\label{iwn1}
I(i\omega_n)&=&  2 \pi c \Sigma \int_{\bf q} \left[{1 \over
{cq^2 + \Sigma}} + {1 \over { cq^2 +dq+\Sigma}} \right. \nonumber \\
& & \left. - {{2(cq^2 +
\omega_n^2 +I(i\omega_n) + \Sigma) + dq} \over {(cq^2 +
\rho_m\omega_n^2+dq + I(i\omega_n) + \Sigma) (cq^2 +
\rho_m\omega_n^2+ I(i\omega_n) + \Sigma)  + \rho_m^2 \omega_n^2
\omega_c^2}} \right]
\end{eqnarray}
\end{widetext}
and the equation determining $\Sigma$ is
\begin{equation} \label{sigmasum}
\Sigma = \sum_K \rho_0^2\Delta_K \frac{{ K}^2}{4\pi}
 e^{- \frac12 { K}^2 B(u_c)}
\end{equation}
The term in the exponential, $B(u_c)$, is given by
\begin{eqnarray}\label{buc}
B(u_c)&=& { 1\over \beta}\int_{\bf q} \left[\sum_{n \neq 0}
\left(G_{cL}(q,\omega_n)+ G_{cT}(q,\omega_n) \right) \right.\nonumber \\
& & \left.+ {1 \over {cq^2 +dq+\Sigma}}+ {1 \over {cq^2 +\Sigma}}\right]
\end{eqnarray}
where the breakpoint  is
\begin{equation}\label{b-uc}
\beta u_c = {{ K^2} \over {8\pi c}}
\end{equation} (determined in Appendix~\ref{ap:solvar}).
Since (\ref{sigmasum})
depends on $I$ through $B(u_c)$,  (\ref{eq:gfn}), (\ref{iwn1}) and
(\ref{sigmasum}) form a closed set of self consistent equations.

\section{Solution of saddle point equations}\label{sec:solution}

We now solve the self consistent equations derived in the previous section.
Before, we proceed with the explicit determination of the solution,
let us examine some of the general features of the solution.
A cursory glance at the propagator (\ref{eq:gfn}), shows that  the mass term $\Sigma$ defines a lengthscale through
$\Sigma=cl^{-2}$.  As explained in Ref.~\onlinecite{giamarchi_vortex_long}, this is the length scale which
separates the Larkin regime  from the random manifold regime.  These two regimes
are characterised by a power law growth of displacements with differing exponents.
Moreover, at zero temperature, in the absence of  thermal fluctuations
the relative displacement  correlation function, $\tilde{B}_{T} (l) \propto { \rm max}[l_c^{2},r_{f}^{2}]$.
Comparing with (\ref{lengths}), we immediately see that $l=R_{c}$ or
equivalently,
\begin{equation}\label{eq:sigmarc}
\Sigma= cR_{c}^{{-2}}
\end{equation}
\noindent
where, $R_{c}$ is
the  Larkin-Ovchinikov length. Contrary to older theories of CDW, $\Sigma$  is
related to $R_{c}$ and   {\it not} the positional
length $R_a$.
We emphasize that  $\Sigma = c R_c^{-2}$ only in the collective pinning regime where $R_c \gg a$. The equivalence $l=R_c$
fails when the system enters the single particle pinning
regime where $R_c <a$.

\subsubsection{Results for $\Sigma$ and $R_{c}$} \label{sec:sig}

To solve for $\Sigma$, it is useful to
 remark  that the structure of the equation (\ref{sigmasum}),
is very similar to the analogous equation for a classical  system with
point-like disorder.
This allows us to directly  apply the formalism developed in Ref.~\onlinecite{giamarchi_vortex_long}
to the present problem.
Re-arranging the terms in (\ref{btbl}), we can see that $B(u_c)$ is
the extent of localization of the particle.
Therefore, to leading order, we replace $B$ in (\ref{sigmasum}) by $l_c^{2}$ which gives
\begin{equation}\label{freq}
\Sigma = c (2\pi^2)^{-{1 \over 6}} R_a^{-2} (a/\xi_0)^6
\end{equation}
where
\begin{equation}
\label{leng}
\xi_0= \text{max}[r_f,l_c]
\end{equation}
The Larkin length $R_c$ is  given by
\begin{equation}
\label{lengthrc}
R_c =  R_a (\frac{\xi_0}{a})^3
\end{equation}
We see from (\ref{freq}), that   $\Sigma$ exhibits very different
magnetic field dependences in the two physically distinct cases of $r_f >< l_c$.
We see that  for $r_{f}\gg l_{c}$, $\Sigma$ is independent of the magnetic field
and for $r_{f}\ll l_{c}$, $\Sigma= bB^{3}$, where reinstating $\hbar$, $b= c (2\pi^2)^{-{1
\over 6}} R_a^{-2} (a^{2}e/\hbar)^{3}$ (at $T=0$).
This has extremely interesting ramifications for experimentally
measurable quantities as we discuss in the following sections.

Let us illustrate the importance of retaining all the harmonics in the case
where disorder varies at length scales smaller than $a$. If we retain only
the lowest harmonic in (\ref{sigmasum}), we would have $r_f=a$ since disorder now varies only on scales
comparable to $a$. In the relevant limit, $l_c < a$,
we  see that the exponential in (\ref{sigmasum}) can be
replaced by unity.
In other words, this corresponds to substituting $\xi=r_f \equiv a$ in (\ref{freq})
leading to the result that $R_c = R_a$.
This precludes the possibility of any dependence of $\Sigma$ on the
magnetic field because $R_a$ is  a static scale independent of $B$.
Thus we see that when disorder varies  at scales smaller than $a$ i.e.,
$r_f < a$,  it is necessary to retain all the harmonics in (\ref{sigmasum})
to obtain the correct result for
$\Sigma$.

\subsubsection{Behavior of $I(\omega)$} \label{sec:iomega}

Though we have solved the equation determining $\Sigma$ in the full quantum
sense, (\ref{iw0}) proves difficult to solve. We thus examine the
semi-classical limit  given by (\ref{iwn1}). As was shown in
simpler cases \cite{giamarchi_columnar_variat}, this limit captures correctly
the main features of the solution.
(\ref{iwn1}) is analytically solvable for small and large frequencies
and can be solved numerically for intermediate frequencies.
Two frequency scales control the behavior of $I(\omega_n)$: (i) the pinning frequency
$\omega_p^{0}={\Sigma}/\rho_m \omega_{c}$, to be discussed in detail in section~\ref{sec:dynamics};
(ii) the cyclotron frequency $\omega_c$.
For small $\omega_n$ ($\omega_n \ll {\frac{\Sigma }{\rho_m\omega_{c}}})$ , one gets
\begin{equation}
I(i\omega_n)= \sqrt {2\rho_m\Sigma
 + {{\pi \rho_m^2 \omega^2_c (c\Sigma)^{1 \over 2}} \over {2 d}}}
\vert \omega_n \vert
\label{iwsmall}
\end{equation}
As expected $I(i\omega_n)\to 0$ when $\omega_n \to 0$. In the limit
of small $\omega_c$, the second term inside the square root can be
nelegected and we obtain the usual result  for a system in a zero
magnetic field. On the other hand, the second term dominates the
low frequency behavior of $I$, if
\begin{equation}\label{limit}
\omega_c^2 \gg
\frac{4d}{\pi\rho_m} \sqrt{\frac{\Sigma}{c}} = \Omega \sqrt{\frac{\Sigma}{\rho_m}}
\end{equation}
This defines a frequency scale, $\Omega^2\sim \frac{d^2}{\rho_m
c}$.  We work in
the limit of large magnetic fields, $\rho_m\omega_c^2 \gg \Sigma$.
Since the two elastic moduli are related $ d \propto
\frac{c}{a}$, $\Omega^2\propto \frac{c}{m^*}$, where $m^*$ is the
effective mass of the system.  This clearly shows that $\Omega$ is
just the plasma frequency of the two dimensional  WC.

For intermediate frequencies $\sqrt{\frac{\Sigma}{\rho_m}} \ll \omega \ll
\frac{\Omega^{2}}{\omega_c}$, the solution becomes
\begin{equation} \label {si1}
I(\omega_n) \simeq {\Sigma \over 6} \log {{{\rho_m^2\omega_n^2 }\omega_c^2}
\over {d \Sigma^{ 3 \over 2}}}
\end{equation}
The analytical solutions  given above, were obtained in the limit of $\rho_m \omega_c^2 \gg \Sigma$.
To obtain the solution numerically, we first  continue (\ref{iwn1})
to real frequencies  $\omega$, which results in a set of two coupled integral equations
for the real and imaginary parts $I_1$ and $I_2$ of $I(\omega)$.
These integral equations can be easily solved numerically.
The numerical solutions indicate that the  linear solution
(\ref{iwsmall}) obtained for frequencies $\omega \ll \Sigma/\rho_m\omega_c$, infact survives beyond this regime.
\begin{figure}
\centerline{\includegraphics[angle=-90,width=\figwidth]{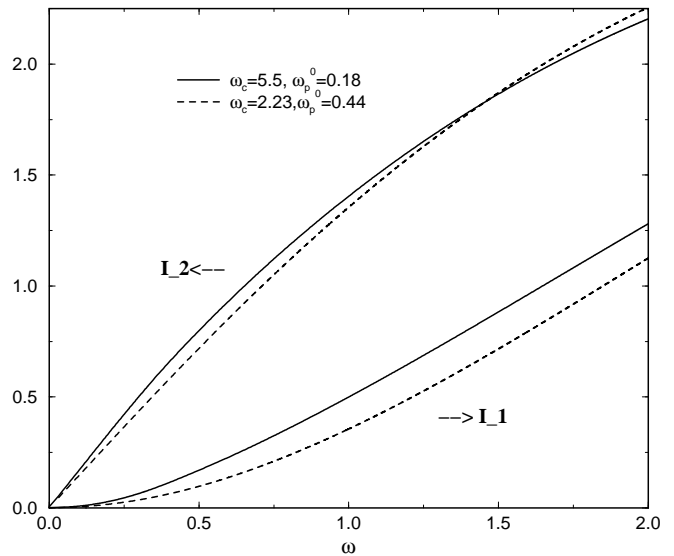}}
\caption{\label{fig:iw} A generic numerical solution  for the real $I_1(\omega)$ and imaginary $I_2(\omega)$
parts of $I(\omega)$ (in units of $\Sigma$) as a function of $\omega$
(in units of $\sqrt{\Sigma/\rho_{m}}$). Results for two different values of $\omega_c$ are shown.
The corresponding value of $\omega_p^0 \equiv \Sigma/\rho_{m}\omega_{c}$ is also given.}
\end{figure}
The numerical approach, however,  is not
constrained to  this regime and can be used to obtain $I$ for all  ranges of $\Sigma$ and $\omega_c$.
Typical plots of $I_1$ and $I_2$ are shown on
Figure~\ref{fig:iw}.
We see from Fig.\ref{fig:iw}, that $I_2$ is indeed linear in $\omega$
and $I_1$ is quadratic for small $\omega$, in agreement of the analytical continuation
of (\ref{iwsmall}).

\section{Physical Properties - Statics} \label{sec:static}

\subsection{Compressibility}
One of the few thermodynamic  properties that can be measured in the QHE
and MOSFET systems is the static compressibility $\kappa$ of the Wigner crystal.
 The formalism used in this paper allows us to calculate the static
compressibility of the disordered crystal.
The calculation of $\kappa$ requires a knowledge of the density-density
correlation function.
\begin{equation}
\kappa= \lim_{q \to 0}\lim_{\omega \to 0} \Pi (q,\omega)
\end{equation}
where the density operator is given by  $\partial_{\alpha}u_{\alpha}= q_{\alpha}u_{\alpha}$.
Clearly,  only the longitudinal components of the displacement
field contribute to the density. Consequently,  the density correlation is entirely
determined by the correlation of the longitudinal displacements and is
given by
\begin{equation}
\Pi(q,\omega)= \rho_0^2 q^{2}\langle u_{L} (q,\omega) u_{L} (-q,-\omega)\rangle
\end{equation}
 First let us consider the pure case. We immediately see that since the coulomb
term $dq$ dominates over the shear term for small $q$, the
compressibility $\kappa=0$  in the pure system. Consequently,
the compressibility remains zero even in the disordered phase.
We reiterate that the compressibility calculated here is
related to the second derivative of the energy with respect to
density  for fixed total number of particles. This also assumes
that the  neutralising background  remains   unchanged as  the
volume of the electron gas is changed.
Before comparing these results with  experiments, it should be ascertained  whether the
compressibility measured in the experiment is the same quantity defined above
\cite{eisenstein_hall_compressibility}.

\subsection{Translational order} \label{sec:translational}

It is interesting to study the effect of disorder on
the translational long range order present in the clean system
at zero temperature. Contrary to
the naive view based on extrapolation of Larkin \cite{larkin_70} and FL
\cite{fukuyama_pinning} methods that
the crystal loses its long range order due to pinning, it was shown
that in $d=3$ the disordered system, although pinned, retains
a defect-free quasi-long range translational order.
In $d=3$, this Bragg glass phase has thus power-law divergent Bragg peaks
\cite{giamarchi_vortex_long,giamarchi_book_young}. For $d=2$
classical systems it is known that topological defects do appear
\footnote{Note that for very smooth disorder $r_f \gg a$
dislocations do not appear at all - even in $d=2$, below a threshold disorder
\cite{carpentier_melting_prl}} but
at a length $R_d$ which is much larger than $R_a$ and thus the
elastic description holds in a large regime of length scales (up to
$R_D$) resulting in a quasi Bragg glass state. The properties of
the $d=2$ quasi Bragg glass for the classical problem have been
studied in detail and the length scale
$R_D$  has been estimated
\cite{giamarchi_vortex_long,carpentier_ledou_triangco,carpentier_melting_prl,ledoussal_dislocations_2d}.

We now derive the result given by the variational method for
the positional correlation functions of the pinned WC in $d=2$.
\begin{widetext}
\begin{eqnarray}
{\tilde B}_{L,T}(x,t=0)&=& {{2 }\over{ \beta u_{c}}}\int\int {{ q dq
d\theta}\over {( 2\pi)^{2}}} d\omega \left[\cos^2 \theta (1-\cos (qx \cos
\theta)){{\Sigma \delta_{\omega=0}} \over {G_{cL,T} (G_{cL,T} + \Sigma)}}
\right.\nonumber \\
   & &+ \left.
(1-\cos^2 \theta) (1-\cos (qx \cos
\theta)){{\Sigma \delta_{\omega=0}} \over {G_{cT,L} (G_{cT,L} +
\Sigma)}}\right]
\label{dispcor}
\end{eqnarray}
\end{widetext}
Since the equal time asymptotic  behavior is governed primarily  by
$\omega=0$,
it is sufficient to retain this mode alone in (\ref{dispcor}). We
find that the  leading contribution to the correlation function grows
logarithmically i.e.,
\begin{equation}
{\tilde B}_{L,T} (x)\simeq { \over {2 \pi c \beta u_c}}  \log
\left({x \sqrt{ \Sigma \over c}}\right)
\label{trcr}
\end{equation}
\noindent
Substituting (\ref{b-uc}) in the above and using the relation
$\Sigma=c R_c^{-2}$
at zero temperature, (\ref{trcr})
simplifies to
\begin{equation}
{\tilde B}_{L,T} (x)\simeq {4\over {{ K}_0^{2}}} \log
\left({x \over {R_{c}}}\right)
\end{equation}
Only the leading contribution to the  displacements grow in an isotropic
manner.
There are non-isotropic corrections to this result but these can be
neglected in the asymptotic limit.
The translational correlation is given by
\begin{equation}
\nonumber
O_T= \langle \exp^{ i{ K}_{0}.{ u}(x)}
 \exp^{- i{ K}_{0}.{ u}(0)} \rangle
\end{equation}
Within the Gaussian approximation used in this paper
\begin{eqnarray}
\nonumber
O_T &=& \exp  [- {{K_{0\alpha}K_{0\beta}} \over 2} {\tilde
B}_{\alpha \beta} (x)] \\
&\simeq&  \left({{R_{c}}\over x}\right)^{2}
\end{eqnarray}
thus the pinned WC also supports quasi long range order.
Since $R_{c}$ is determined by (\ref{lengthrc}), we see from (\ref{trcr})
that the growth of displacements will have a magnetic field dependence
depending on whether
$r_f $ is lesser or greater than $l_c$. The $B$ dependence in
a static correlation function like the one studied here is
a direct consequence of the fact that in the quantum system the statics and
the dynamics are
coupled.

Although the GVM correctly captures several features of the
growth of displacements in $d=2$, it should be taken with a grain
of salt concerning the exact asymptotic behavior. It is well known for the
classical problem that the variational method slightly underestimates
the fluctuations at the lower critical dimension $d=2$
and that the variational result $\log(r)$ should in fact
be replaced by $\log^2(r)$. The latter result can be derived by a
more accurate renormalization group approach
\cite{cardy_desordre_rg,giamarchi_vortex_long,carpentier_ledou_triangco,giamarchi_book_young}.
It is fair to expect that
the result will be similar for the quantum problem considered here,
since  the $\omega=0$ mode considered in
(\ref{dispcor})  mimics
a $d=2$ classical problem. Despite this,
even a $\log^2$ growth of the correlation functions
would lead to an
extremely weak destruction of the positional order, in the quantum case analogous
to the quasi Bragg glass. The other important question is the
role of dislocations.
Again the precise answer is only known for the classical problem, for which
it has been shown that in $d=2$ dislocation are  generated {\it but} at a lengthscale $R_D$
much larger \cite{giamarchi_vortex_long} than $R_a$.
In particular the result at zero temperature, relevant for the ground state, reads
\cite{ledoussal_dislocations_2d,carpentier_traveling_xy}:
\begin{equation} \label{eq:rd}
R_D \sim R_a e^{ c \sqrt{(\frac{1}{8} -  \sigma_0) \ln (R_a/a)} }
\end{equation}
where $\sigma_0$ is proportional to the strength of disorder.
For the pinned WC the effect of additional
quantum fluctuations and Coulomb interactions remains to be treated
in a precise way. As can be seen e.g. by  analogy to a
3d classical system with correlated disorder (where it is known that fluctuations
play a minor role), one can surmise that (\ref{eq:rd}) should be rather stable
to small quantum fluctuations. Consequently,
the pinned WC should still exhibit an extremely good positional order for weak
disorder.

\section{Physical Properties - Dynamics} \label{sec:dynamics}

\subsection {General features of the conductivity} \label{sec:generalcond}

The formalism and the results obtained in the previous sections
can now be used to calculate the magneto-conductivities of the
pinned crystal. The conductivities are given by
\begin{equation}
\sigma_{\alpha \beta}(\omega)= i \rho_{c}^{{2}}\omega G_{\alpha \beta} (q=0,
\omega+ i\epsilon)
\end{equation}
$\alpha,\beta= x,y$
and the $G$ are the  displacement Green's function.
The real frequencies $\omega$ are analytic continuations of the  Matsubara frequencies $\omega_n$.
Using (\ref{gc}), the
explicit form of the conductivities are
\begin{eqnarray}
\sigma_{xx}& = \sigma_{yy}&=\rho_{c}^{{2}} {{i\omega [ -\rho_{m}\omega^2 + \Sigma + I
(\omega)]} \over { (\Sigma - \rho_{m}\omega^2 + I ( \omega))^2 - \rho_m^2
\omega_c^2 \omega^2}}
\nonumber \\
\sigma_{xy}&= -\sigma_{yx}& =  \rho_{c}^{2}{{\omega [\rho_m \omega \omega_c]}
\over { (\Sigma - \rho_{m}\omega^2 + I ( \omega))^2 - \rho_m^2\omega_c^2
\omega^2}} \label{eq:cond}
\end{eqnarray}
The conductivity tensor  is  therefore, completely determined by
$\Sigma$ and $I(\omega)$. As usual the Hall resistivity is given by
\begin{equation}
R_H= {{\sigma_{xy}} \over {\sigma_{xx}^2 + \sigma_{xy}^2}}
\end{equation}

For the pure system, $\Sigma=I=0$ and (\ref{eq:cond}) gives back the
standard features of electrons in a magnetic field. The Drude peak is
shifted by the magnetic field from $\omega=0$ to the cyclotron frequency $\omega_c$.
The d.c. Hall conductivity has a finite value
$\sigma_{xy}(\omega=0)=\rho_c/B$, and the Hall resistivity has the classical value $R_H = B/\rho_c$.

In the presence of disorder ($\Sigma\ne 0$, $I(\omega)\ne 0$),
as can be seen from (\ref{eq:cond}), the $\delta$-function peak at $\omega_c$
will broaden and slightly shift in frequency. (\ref{eq:cond}) also shows that the dc conductivities
$\sigma_{xx}(\omega=0)=\sigma_{yy}(\omega=0)=\sigma_{xy}(\omega=0)=0$,
both in the presence and absence of a magnetic field. This  is a consequence of
 the crystal  being  pinned by  disorder.  More importantly,
a new peak appears in the diagonal conductivity
at new scale $\omega_p$ called  the pinning
frequency.

A simple estimate of the pinning frequency is given by the poles in the
conductivity when setting $I(\omega)=0$.  This gives a $\omega_{p}^{0}=
(\sqrt{\omega_{c}^{2}+ 4 \Sigma/\rho_m}- \omega_{c})/2$. In the limit of large magnetic fields,
which is relevant for the case of quantum Hall systems, the above expression
simplifies to
\begin{equation}\label{omega_pzero}
\omega_p^0= {\Sigma \over {\rho_m\omega_c}}
\end{equation}
The cyclotron
peak is shifted to $\omega_{c}+\omega_p^0$.
Of course $\omega_p^0$ is only an approximation to the true pinning
frequency $\omega_p$ that should be determined directly by the maximum of the conductivity
(\ref{eq:cond}). The main difference between $\omega_p^0$ and $\omega_p$ stems  from the presence
of   $I(\omega)$  in the denominator.
\begin{figure}
\centerline{\includegraphics[angle=-90,width=\figwidth]{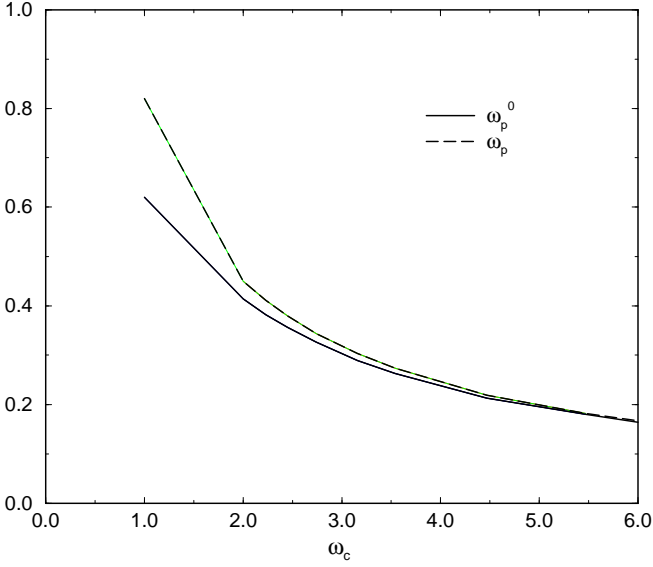}}
\caption{\label{fig:fwpwp0} $r_f > l_c$: The pinning frequencies $\omega_p$ and $\omega_p^0$
as a function of  $\omega_{c}$. All  frequencies are in  units of  $\sqrt{\Sigma/\rho_m}$.
Note that $\omega_p^0$ and $\omega_{p}$  become quite different for low magnetic fields.}
\end{figure}
\begin{figure}
\centerline{\includegraphics[angle=-90,width=\figwidth]{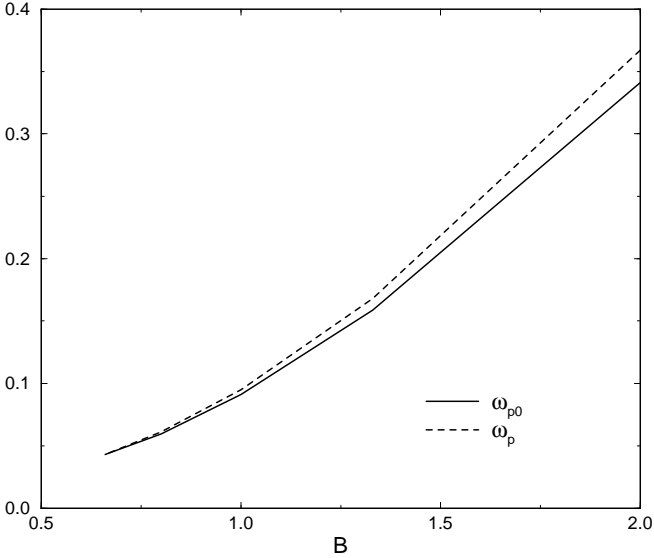}}
\caption{\label{fig:bwpwp0}
$r_f < l_c$: The pinning frequencies $\omega_p$ and $\omega_p^0$
(in units of $\rho_c^3/ \rho_m^2 b$)
as a function of the applied field $B$ (in units of $\rho_c^2 / \rho_m b$). $b$ is defined in Sec.~\ref{sec:sig}.}
\end{figure}
Typical values of the pinning frequencies are shown in Fig.~\ref{fig:fwpwp0} and  Fig.~\ref{fig:bwpwp0}.

Quite remarkably, the structure (\ref{eq:cond}) of the
conductivity tensor shows that the Hall resistivity $R_H$ is a
constant independent of $\omega$
\begin{equation}
R_H= \frac{B}{\rho_c}
\end{equation}
as in the pure system. This is a remarkable result since since it shows that the disorder has
no influence on the Hall resistivity of a pinned WC. This property comes from the
fact that the disorder is local, which in the variational calculation makes all off diagonal self
energies zero, as explained in Appendix~\ref{ap:solvar}. Since this feature is linked to the
locality of disorder it is expected to be valid beyond the variational approximation.

\begin{figure}
\centerline{\includegraphics[angle=-90,width=\figwidth]{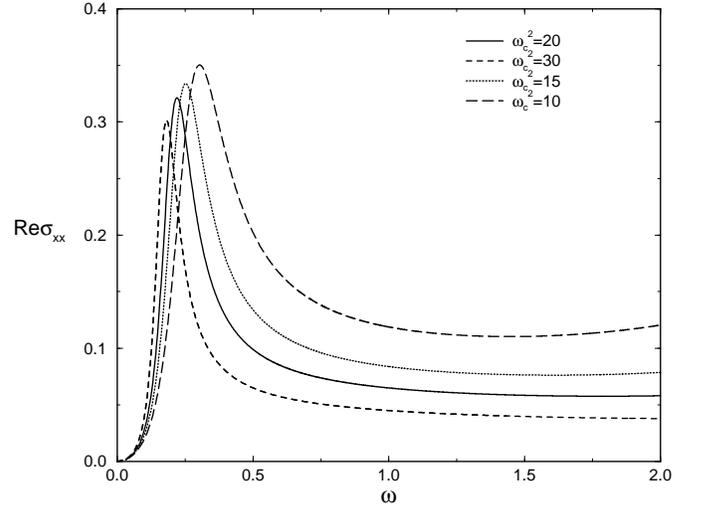}}
\caption{\label{fig:fcond}
$r_f> l_c$: Re$\sigma_{xx}$ (in units of $\rho_c^2 /\sqrt{\rho_m
\Sigma}$) as a function of $\omega$ and $\omega_{c}$(in units of $\sqrt{\Sigma / \rho_m}$).
Note the non Lorentzian shape of the peaks. In this regime, the pinning frequency decreases with
increasing magnetic field.}
\end{figure}

\begin{figure}
\centerline{\includegraphics[angle=-90,width=\figwidth]{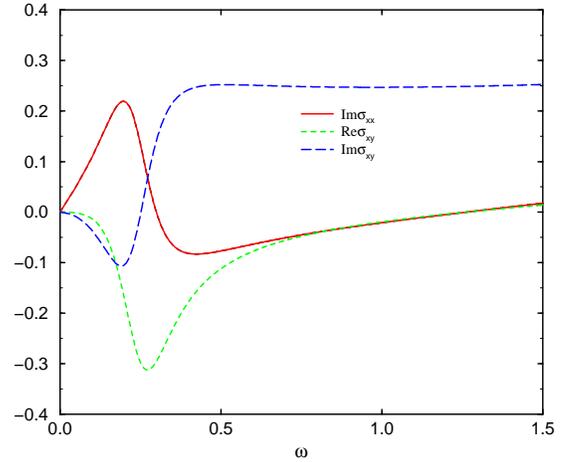}}
\caption{\label{fig:Imcond}
$r_f > l_c$: Im$\sigma_{xx}$,Re$\sigma_{xy}$ and Im$\sigma_{xy}$(in units of $\rho_c^2 /\sqrt{\rho_m \Sigma}$)
as a function of $\omega$ and $\omega_{c}$ (in units of ${\sqrt {\Sigma / \rho_m}}$)
for $\omega_{c}^{2}= 10$.}
\end{figure}

We now analyze the frequency dependence of the conductivities.
For small values of $\omega$, substituting (\ref{iwsmall}) in (\ref{eq:cond}) gives
\begin{eqnarray}
Re\sigma_{xx} ( \omega)&=&
 \rho_c^2{\sqrt {2\rho_m\Sigma
 + {{\pi \rho_m^2\omega^2_c (c\Sigma)^{1 \over 2}} \over {2d}}} {{\omega^2}
\over {\Sigma^2}}}\nonumber \\
Re\sigma_{xy} ( \omega)&=&
  \rho_c^2 \rho_m\omega_c {({\omega \over \Sigma})^2}
\end{eqnarray}
\noindent
and the imaginary or dissipative part of the conductivity grows as
\begin{eqnarray}
Im\sigma_{xx}(\omega) &=&
 \rho_c^2 {\omega \over \Sigma}  \nonumber \\  \
 Im\sigma_{xy}(\omega)&=&
\rho_c^2 \rho_m^{3 \over2} {{\omega_c
\omega^3} \over { \Sigma^{ 5 \over 2}}}    \label{cxx}
\end{eqnarray}
In the region
$ \sqrt{\frac{\Sigma}\rho_{m}} \ll \omega \ll \frac{\Omega^{2}}{\omega_c}$
we find using  (\ref{si1})
\begin{equation} \label{cxx1}
Re\sigma_{xx}(\omega)  \sim {\rho_c^2 \over \rho_m^2} {\Sigma \over
{\omega_c^2 \omega}} \qquad
Re \sigma_{xy}(\omega)  \sim {\rho_c\over B}
\end{equation}
\noindent
The full behavior of the conductivity can
be obtained from (\ref{eq:cond}) by numerically solving the equation for $I(\omega)$.
Typical plots of diagonal and transverse
conductivities are shown in Fig.~\ref{fig:fcond}
and Fig.~\ref{fig:Imcond}.

In the limit of very strong magnetic fields, since $\omega_p^0 \ll \sqrt{\frac{\Sigma}{\rho_m}} \ll \omega_c$, provided
$\omega_c^2 \gg \Omega \sqrt{\frac{\Sigma}{\rho_m}}$ (see discussion in Sec.~\ref{sec:iomega})
the evolution of the width $\Delta \omega_p$ of the pinning peak
is completely governed by the low frequency
behavior of $I$ obtained in (\ref{iwsmall}). In this limit,
\begin{equation}\label{width}
\frac {\Delta \omega_p} {\omega_p} \propto \left(\frac{\Sigma}{n}\right)^{1/4}
\end{equation}
Again from (\ref{freq}), we see that the behavior of the peak widths is again dictated by
$r_f$ and $l_c$. A similar result was obtained in Ref.~\onlinecite{fogler_pinning_wigner} with an
undetermined exponent $s$ (to  be discussed in Sec.~\ref{sec:otherth}).

\begin{figure}
\centerline{\includegraphics[angle=-90,width=\figwidth]{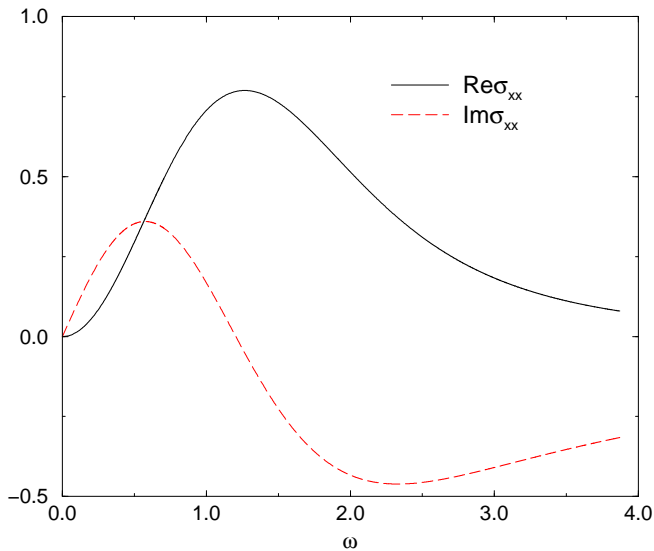}}
\caption{\label{fig:cdw}
Re$\sigma_{xx}$ and Im$\sigma_{xx}$ (in units of $\rho_c^2 /\sqrt{\rho_m \Sigma}$)
as a function of $\omega$(in units of ${\sqrt {\Sigma / \rho_m}}$) for the
pinned crystal in zero magnetic field.}
\end{figure}

The case of zero  external
magnetic field can also be obtained from (\ref{eq:cond}) by setting $\omega_c=0$.
It leads to different energy scales and physical behavior.
The clean crystal has infinite dc conductivity and zero
ac conductivity. In the presence of disorder, the dc conductivity is
zero as expected and a pinning peak develops at the frequency
$\omega_p^0 = {\sqrt {\Sigma/\rho_m}}$ where $\Sigma$ is determined
by (\ref{freq}) with $\xi=r_f$. We show a typical plot of the conductivity
in Fig.~\ref{fig:cdw}.

An important feature of our results is that
the pinning peaks are rather sharp for finite magnetic fields, as opposed to the
rather broad pinning peak obtained in the limit of zero magnetic field.
This is expected because the cyclotron peak in the former case
has appreciable spectral weight.
Comparing Fig.~\ref{fig:cdw} and Fig.~\ref{fig:fcond}, we see that
the peaks in the finite field case are at least a factor of 10-15  times
narrower.

Moreover, the older theories predict
a Lorentzian broadening of the peaks. A glance at the conductivities
plotted in Figs.\ref{fig:fcond},\ref{fig:bcond} and \ref{fig:cdw}   shows that the pinning
peaks are  non-Lorentzian and asymmetric around
$\omega_p$. This is an important prediction of our theory. This asymmetry in the
peak is indeed seen in microwave conductivity measurements on
certain Ga-As samples
\cite{li_conductivity_wigner_magneticfield,li_conductivity_wigner_density,beya_thesis}.
Comparison of our results with experiments will be discussed in further detail in Sec.~\ref{sec:experiments}.

A generic property  of pinned systems is that
a finite nonzero {\bf threshold electric field $E_T$} is required before
the crystal de-pins and d.c. current flows through the system.
Since $E_T$ is intrinsically the signal of non-linear response, it cannot
be directly evaluated using Kubo-like formulae. As for the dynamics of
classical crystals \cite{chauve_creep_short,chauve_creep_long}, its precise calculation would
require a non equilibrium formulation such as the Keldysh technique.
This is clearly a formidable task, but fortunately for the classical
cases, approximate derivations allow one to relate $E_T$ to equilibrium quantities.
A dynamical derivation has shown that such formulas indeed yield the
correct threshold field $E_T$. Here, we use the same approximate
approach to derive $E_T$ for the quantum system.
 In the  collective pinning
 regime, $R_c \gg a$, the threshold  field $E_T$
 is determined by $R_c$ \cite{larkin_ovchinnikov_pinning}
\begin{equation}\label{eq:thresh}
E_T= {c } R_c^{-2} \xi_0
\end{equation}
\noindent
where $\xi_{0}= \rm {max}[r_{f},l_{c}]$.
Clearly,   $E_T$ defined by (\ref{eq:thresh}) is directly linked to $\omega_p^0$ by
$E_T=\rho_m\omega_c \omega_p^0 \xi_0$.
The validity of this result remains to be explicitly proven in the
quantum case.

\subsection{Magnetic Field dependence}

As already pointed out, it is crucial to  have theoretical predictions for
the various features  and attributes of the conductivity as a function of physical
parameters like the magnetic field and the density.
This would then facilitate  direct comparisons between
experiment and theory  and consequently further our understanding of the
system.

It is clear from the previous section that $\Sigma$ and hence the length scale
$R_c$ determine most of the
dynamical properties of the pinned crystal. Different behaviors
are possible depending on the relative magnitude of the correlation length
of the disorder $r_f$ and the particle size
$l_c$ (see Fig.~\ref{fig:basiclength}). This leads to a
physics very different from that predicted by the naive weak pinning
scenario \cite{fukuyama_pinning,normand_millis_wigner} where, it is implicitly
assumed that the scale $R_a$ is the only lengthscale in the system.
Two main behaviors can be obtained.

\subsubsection{$r_f >l_c$} \label{sec:simple}

In this case, the particle  locally sees a smooth potential, and thus
behaves effectively like a point particle. Hence the length $l_c$ is
not directly relevant. This leads to a $R_c$
which is independent of the magnetic field, as was shown in Sec.~\ref{sec:solution}.
>From (\ref{freq}), one obtains for the naive pinning frequency
\begin{equation} \label{eq:naivecdw}
\omega_p^{0}(B)= {{\Sigma} \over {\rho_{m} \omega_{c}}}=  (2\pi^2)^{-{1
\over 6}} {{c R_a^{-2}}\over {\rho_{c}B}}({a \over {r_{f}}})^6
\end{equation}
Consequently,
the naive pinning peak moves towards the origin as $B^{-1}$ and gets narrower
with increasing field. However, as mentioned in Sec.~\ref{sec:generalcond},
the true pinning frequency is given by the location of the peak in the
optical conductivity, and can be in principle different from $\omega_p^0$.
A plot of the conductivity
in this regime for different values of the magnetic field is shown in
Fig.~\ref{fig:fcond}. From it we extract the true pinning frequency $\omega_p$.
As seen in Fig.~\ref{fig:fwpwp0}, both frequencies decrease as $B^{-1}$.
We see that
the difference $\omega_p -\omega_p^0$  is more pronounced for small values of
$B$ or equivalently, $\omega_{c}^{2}/\Sigma$, and goes to zero for
increasing values of the field $B$. Similarly, the non-Lorentzian nature of the peaks discussed earlier  is
enhanced for small and intermediate magnetic fields (cf. Fig.~\ref{fig:fcond}).
The regime $r_f > l_c$ includes the case of the conventional CDW
in a magnetic field for which $r_f \sim a$
and thus $R_a \simeq R_c$. Therefore, CDW have the same magnetic field
dependence for the pinning frequency as the Wigner crystal in this regime.
The density dependences are,  however,  quite different, as will be shown
in Secs.~\ref{sec:density} and \ref{sec:experiments}.

\subsubsection{$r_f < l_c$}

In this regime, since the size of the particle is larger than the correlation
length of the disorder (see Fig.~\ref{fig:basiclength}), the particle locally
averages over the disorder. The effective disorder seen by the particle thus
strongly depends on its size $l_c$. Since the particle size $l_c$ is strongly
magnetic field dependent, this regime exhibits anomalous magnetic field dependence
in all physical quantities. This physics is captured by the equations for
$\Sigma$ (\ref{sigmasum}). (\ref{freq}) indicates that
$\Sigma$ is strongly $B$ dependent.
This  explicit magnetic field dependence
results in a naive pinning frequency (we have reinstated $\hbar $)
\begin{equation} \label{eq:naivewc}
\omega_p^{0}(B)= (2\pi^2)^{-{1 \over 6}} {{c R_{a}^{-2}B^{2}} \over {\rho_{c}}}({{ e a^{2}} \over
\hbar })^3
\end{equation}
which increases quadratically with increasing field $B$.
Therefore, unlike the previous case (\ref{eq:naivecdw}) where the
pinning frequency and peak width were decreasing functions of $B$, here
the pinning peak moves up and broadens with increasing field.

\begin{figure}
\centerline{\includegraphics[angle=-90,width=\figwidth]{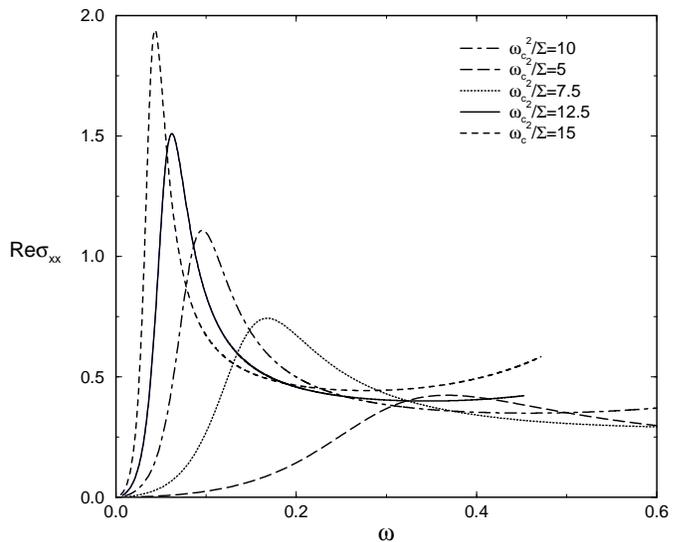}}
\caption{\label{fig:bcond} $r_f < l_c$: Re$\sigma_{xx}$(in units of $b\rho_m /\rho_c$)
 as a function of $\omega$ (in units of $\rho_c^3 / \rho_m^2 b$; $b$ has been
defined in the text). Note that increasing $\omega_{c}^{2}/\Sigma$
corresponds to decreasing magnetic fields.}
\end{figure}

As was done in  Sec.~\ref{sec:simple}, it is important to
extract the true pinning frequency and peak width from  the  full
 conductivity. $Re\sigma_{xx}$ has been plotted
for various values of  $B$ ranging intermediate fields to strong fields, in  Fig.~\ref{fig:bcond}.
In this regime, all curves are plotted for varying
$\omega_{c}^{2}/\Sigma$. However, since $\Sigma \propto B^{3}$,
$\omega_{c}^{2}/\Sigma \propto  B^{-1}$, which implies that an
increase in  $\omega_{c}^{2}/\Sigma$ corresponds to a decrease in the
magnetic field $B$.
A plot of the true pinning frequency is shown in Fig.~\ref{fig:bwpwp0}.
One sees that in this case the shift in pinning frequency, compared to the naive
expression (\ref{eq:naivewc}) is quite important. As before,
this difference is
more pronounced for  small and intermediate values of
$\omega_{c}^{2}/\Sigma$ which contrary to the previous case, now corresponds to large values of the field $B$.  In a similar vein,  as seen in
Fig.(\ref{fig:bcond}),  the peaks become
more and more non-Lorentzian for increasing $B$ i.e., decreasing $\omega_{c}^{2}/\Sigma$.
Our result for the field dependence of $\omega_p$, shown in Fig.~\ref{fig:bwpwp0},
can be  compared with the experimental results. This point will be discussed
in further details in Sec.~\ref{sec:experiments}.

As long as $r_f < l_c$, $\omega_p^0$ increases  with $B$.
This increase in the pinning frequency is not
indefinite. Two effects can limit this regime. For sufficiently large $B$,
since $l_c$ decreases with $B$ one will cross over to the regime of
Sec.~\ref{sec:simple} for which $l_c < r_f$. Since from (\ref{freq}) and
(\ref{eq:sigmarc}), $R_c$ decreases with increasing $B$, this regime is also limited
by $R_c \simeq a$, below which the system is no longer in the collective pinning regime.
In that case, the correspondence (\ref{eq:sigmarc}) between $R_c$ and  $\Sigma$ no longer holds
and the result (\ref{freq}) for $\Sigma$ is invalid. One should thus re-evaluate
$\Sigma$ in this single particle pinning regime. Solving
(\ref{sigmasum}) in this regime, we find $\Sigma \propto B^{3 \over 2}$ and
\begin{equation}
\omega_p^{{0}} \propto B^{1\over 2}
\end{equation}
Which of the two crossovers ($l_c \sim r_f$ or $R_c \sim a$) occurs first depends on the
strength of the disorder in the system considered.

\subsubsection{Summary of magnetic field dependence} \label{sec:summary}

In general, one can thus expect three regimes with  dramatically different
behaviors of  experimentally measurable quantities. A summary of these regimes
and their features is shown in Table~\ref{tab:table}.
Contrary to the standard CDW case, for a Wigner crystal the pinning frequency
can increase with the magnetic field provided one is in the regime $r_f < l_c$.
In all the regimes, the height of the pinning peak
decreases approximately as $B^{-1}$  with increasing field.
As expected, the cyclotron peak at $\omega_c + \omega_p$
is not qualitatively affected by the relative magnitudes
of $r_f$ and $l_c$ and always moves upwards with increasing $B$.

The difference between $\omega_p^0$ and $\omega_p$ that appears
in the WC regime, urges a  re-examination  of the relation between the
pinning frequency and the threshold field. Indeed if the formula (\ref{eq:thresh})
holds, the threshold field should be related to $\omega_p^0$ and not $\omega_p$.
This would imply that in the WC regime, the  pinning frequency and threshold field
would not be linearly related. It would thus be interesting to have extended plots
of $E_T$ vs the pinning frequency $\omega_p$, such as the ones of Ref.~\onlinecite{williams_wigner_threshold}.
Let us reiterate  that
although the pinning frequency $\omega_p$ can be systematically calculated, the
expression (\ref{eq:thresh}) for the threshold field remains unverified in the
quantum case. Assuming (\ref{eq:thresh}), the magnetic field dependence of the the threshold field $E_T$ and the
dielectric constant  defined by
\begin{equation}
\epsilon_{xx}= Im\sigma_{xx}/\omega
\label{eq:dielectric}
\end{equation}
 are summarized in Table.~\ref{tab:table}.
\begin{table}
\begin{tabular}{lcccc}
\colrule
Regime  &  $\Sigma$ &  $\omega_p^0$ &  $E_T$ & $\epsilon_{xx}$ \\
\colrule
$r_f > l_c$  &  $B^0$ &  $B^{-1}$ &  $B^0$ &  $B^0$ \\
$r_f < l_c$  &  $B^3$ &  $B^{ 2}$  &  $B^{5 \over 2}$ &  $B^{-3}$\\
$R_c < a$  &  $B^{3 \over2}$  &  $B^{ 1 \over 2}$ &  $B$
& $B^{-{3 \over 2}}$ \\
 $n$ ($r_f >< l_c$) &  $n^{-\frac1{2}}$ &  $n^{-\frac3{2}}$ & $n^{-\frac1{2}}$&
$n^{\frac5{2}}$ \\
\colrule
\end{tabular}
\caption{Magnetic field and density dependences of various dynamical quantities.
\label{tab:table}}
\end{table}

\subsection{Density dependence}\label{sec:density}

Another way to probe the Wigner crystalline state is to fix the magnetic field and
see how the pinning peak evolves as the density of electrons is varied slowly.
The first thing to realize is that since $r_f$ and $l_c$ are both independent of the
density, the density dependence of $\Sigma$ in (\ref{freq}) and hence
$\omega_p^0$ are the same  irrespective of the
relative magnitudes of $r_f$ and $l_c$.  The  pinning  frequency using (\ref{freq})
is given by
\begin{equation}
\omega_p^0= {\pi \over {m\omega_c}}  {{\epsilon_0 \Delta} \over
{ e^2 \xi_0^6}} {a}^{3} \propto n^{-{3 \over 2}}
\label{eq:wpdens}
\end{equation}
\noindent
(\ref{eq:wpdens}) shows that the pinning frequency increases as density
is decreased. The width of the peak increases and the height of the
pinning peak decreases as $n^2$ with decreasing $n$. This is analogous
 to
the behavior seen for increasing $B$ when $r_f < l_c$.
Accordingly,  other dynamical quantities like $E_T$, $\epsilon_{xx}$
etc have the density dependences indicated in Table.~\ref{tab:table}.

\subsection{Limitations and dc conductivity}  \label{sec:limitations}

Before embarking on a comparison of our results with
experiments, let us discuss the range of validity of our theory.

As already mentioned in Sec.~\ref{sec:elasticmodel}, the
quantum nature of the original particle can be safely hidden in
the parameters of the crystal hamiltonian, as long as the spatial extent of
the particles remain small compared to the lattice spacing.
 One expects this to be valid even
quite close to the melting transition of the WC.
Of course, our approach does not allow for a very
precise description of the melting properties, beyond the simple
position of the transition, since this simplified description is incapable of
taking into account the fermionic statistics and
the incompressible quantum fluid that is the melted ``crystal''.

An a priori limitation of our study, even in the
insulating phase far from melting, is to know whether the elastic model is
a satisfactory starting point to obtain the transport properties
or whether defects (vacancies, interstitials, dislocations etc)
should be taken into account from the start. Indeed if dislocations
appear at a typical lengthscale $R_D$ they
will strongly modify the results given by the elastic
description for frequencies smaller than a characteristic frequency
$\omega_D \propto R_D^{-1}$.

\begin{figure}
\centerline{\includegraphics[width=\figwidth]{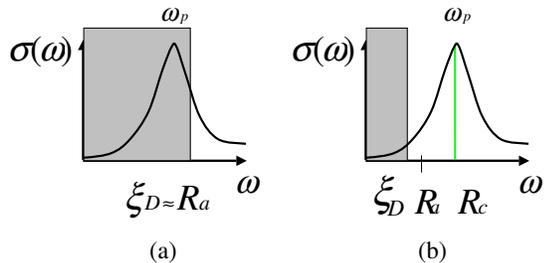}}
\caption{\label{fig:disloc} (a) If dislocations  occurred at scale $R_a$ and the pinning frequency was
controlled by $R_a$, as was naively believed, the elastic theory is incapable of giving any  reliable information on the
pinning peak. It would be necessary to include dislocations from the start. (b) As was shown in the text, dislocations
occur in fact at $R_D \gg R_a$ and the pinning peak depends on $R_c \ll R_a$. Thus the pinning peak is given
quantitatively by a purely elastic theory. It is necessary to take into account topological defects such as
dislocations only at much lower frequencies, and in particular if one wants to obtain reliable results for the d.c. transport.}
\end{figure}

Older theories of disordered elastic systems suggest that due to the
presence of disorder, defects are favored at the length $R_a$ for
which displacements would become of order of the lattice spacing
(see section~\ref{sec:lengthscales}).
If, as in FL description, the pinning peak would be associated
with the length $R_a$, we see that dislocations would  change
most of the interesting optical conductivity and in particular
the pinning peak. Such a situation is summarized in
Figure~\ref{fig:disloc}(a).
If this was correct this
would seriously impair any description of the transport
properties based on an elastic hamiltonian.

In fact, both of these commonly accepted points are {\it incorrect},
which makes the description based on an elastic model much {\it more useful} than
initially  anticipated. Firstly, as we already discussed in
Sec.~\ref{sec:generalcond}, the pinning peak is not associated with the length
$R_a$ but with the a priori much smaller lengthscale $R_c$ (note that
larger lengthscales correspond to smaller frequencies). Secondly,
as discussed in Section~\ref{sec:translational} defects in disordered elastic systems
are much less important than initially believed and appear at a lengthscale $R_D$
much larger than $R_a$ ($R_D=\infty$ in $d=3$ for weak disorder).
Since $R_D > R_a > R_c$, the frequency
$\omega_D$ is thus much smaller than the characteristic frequencies of the pinning peak $\omega_p$.
This implies that dislocations will not drastically affect the pinning peak and thus
that the whole pinning peak structure is given {\bf quantitatively} by the elastic
model. This is summarized on Figure~\ref{fig:disloc}(b). Our study is thus perfectly adapted to
describe optical conductivity experiments.

For d.c. transport or static properties the role played by topological defects is less obvious.
In particular, it is reasonable to assume that they can have an impact on some of the static properties.
Their relevance for  the thermodynamics is hard to estimate but very likely to be reasonably
small since the contribution of the elastic part of the hamiltonian is usually finite and thus dominant.
A notable exception might be the compressibility. However, dislocations are mostly expected to play a role in the
d.c. transport properties. Indeed a pinned system, in the absence of topological defects  has a zero d.c. linear
response to an external force, as was shown in Sec.~\ref{sec:generalcond}. The calculation of the nonlinear
response has only been done for classical systems for which it has been shown that to move, the system has
to de-pin collectively larger and larger bundles, as the external force $F$
becomes small, leading to a very small velocity of the form $v \sim e^{-\beta(1/F)^\mu}$.
For a quantum elastic system,  it is yet unclear how this creep law should be modified \cite{blatter_vortex_review},
and what would be
the d.c. transport of the purely elastic system. Nonetheless, the existence of topological defects
allows for a linear, even if exponentially small, response
to exist, and this should dominate the d.c. linear response.
The temperature dependence of the d.c. resistivity is also clearly linked to the nature and mobility
of the defects in the crystal. Clearly, obtaining the dynamics of a disordered system in the presence of
topological defects is a formidable task.
This again emphasizes the importance of optical conductivity experiments to analyse systems expected
to be pinned elastic structures, since they permit a comparison with controlled theories.

\section{Experiments} \label{sec:experiments}

Several experiments measured the optical properties of a two dimensional electron gas subjected to
a strong magnetic field. One important question is of course whether these
data, obtained in the regime of small filling fractions where crystallization is expected, can
be satisfactorily explained within a pinned WC theory, or whether more exotic descriptions are necessary.

A first batch of experiments \cite{li_conductivity_wigner_magneticfield,hennigan_optical_wigner,beya_thesis}
obtained the pinning frequency and the general frequency profile of the
pinning peak for various values of the applied magnetic field. Important features of these experiments
were a pinning peak whose frequency increased with increasing magnetic field.
This qualitative behavior is in contradiction with the naive expectation based on the Fukuyama-Lee
arguments\cite{fukuyama_pinning}, that the pinning frequency decreases with increasing $B$. This was interpreted
as casting doubts on whether such peaks could be interpreted within a pinned WC phase.
As we pointed out, this conclusion is incorrect, and the experiments are in fact consistent, within
our theory, with a pinned WC. Indeed, as we pointed out the  FL-type theories incorrectly assume that
$R_a$ is the lengthscale controlling the pinning peak, while the relevant lengthscale is $R_c$.

As was discussed in Sec.~\ref{sec:iomega}, the detailed dynamical properties depend on the disorder and  on whether
$\omega_{c}^{2} \gg  \Omega \sqrt{\Sigma/\rho_{m}}$ or not.
The typical experimental parameters  for the hole-like samples are: $B\sim 13T$,
$\omega_c= 7.6 \times  10^{12} Hz$,
$m^*=0.3 m_e$, where $m_e$ is the bare electron mass, $\epsilon= 13 \epsilon_0$ ($\epsilon_0$ is the
dielectric constant of the vacuum) and densities of the order of $n=5\times 10^{10}cm^{-2}$.
For these values, using the classical values of the elastic moduli, the plasma frequency
$\Omega = 4\times  10^{12} Hz$. Since the measured pinning frequencies  are of the order
of $\omega_p \simeq 7 \times  10^{9} Hz$, we see that the condition (\ref{limit}) is easily satisfied,
implying that the experimental systems lie  in the  strong field regime discussed in Sec.~\ref{sec:generalcond}.
Here, the pinning frequency can in fact increase
with the magnetic field \cite{chitra_wigner_hall} (cf. Sec.~\ref{sec:generalcond}).
Thus the observed behavior is in fact compatible with the one of a pinned WC.
Our theory also predicts a non lorentzian shape for the peaks, as shown in Fig.~\ref{fig:bcond},
a feature that is clearly exhibited by the data \cite{engel_conductivity_lorentzian}.

However several features seen in these experiments are still extremely puzzling, and in fact incompatible,
to the best of our knowledge,
with all existing theories. Some anomalous features are the saturation of the pinning frequency and the
increase of spectral weight in the pinning peak when the magnetic field in increased.
Within a WC approach, if one in addition assumes field independent elastic constants,
one always expects a reduction of the spectral weight contained
in the pinning peak due to the continuous transfer of spectral weight from the pinning peak at low
frequencies  to the cyclotron peak. On the other hand, it is crucial to note
 that the interpretation of the
experimental data is complicated by the following facts: (i) the experiment is performed at a quite high temperature
$T=50mK$ which is comparable with the pinning frequency of $f\sim 1GHz$. Therefore,
thermal contributions to the broadening of the peaks might have to be taken in. (ii) The filling factors considered are quite high.
It is thus plausible that most of the data lies outside the range of validity of a putative WC.
It would thus be extremely useful, to have more detailed experiments performed in a lower range of temperatures
and at lower filling fractions. (iii) Since the experiment is performed at relatively large filling factors,
the very hypothesis of field independent elastic constants, which is valid at large fields, is questionable.
One would have to know the true dependence of the elastic constants on the magnetic field, to make
a reliable comparison even within the framework of a pinned WC theory.

Another series of experiments \cite{li_conductivity_wigner_density} focused on the density dependence of the pinning peak.
This set of experiments
were done at lower temperatures $T=25mK$ and lower filling fractions, so this data is much more likely
to be in a regime of a WC. Since these experiments are at fixed magnetic field, they partially avoid the above mentioned
problem with the elastic constants. They should thus provide for a more clear cut interpretation.
A decrease of the pinning frequency with increasing density was observed, fitting reasonably well with
either a $n^{-\frac3{2}}$ or a $n^{-\frac1{2}}$ behavior, depending on the sample. A comparison of these
results was made with the FL theory for CDW. Within this theory {\it and  for a CDW}, the pinning frequency
varies as $n^{-\frac1{2}}$. A blind application of this result
to the WC led Ref.~\onlinecite{li_conductivity_wigner_density} to conclude
that there was  good agreement between the
theory and the experimental findings. But  as shown in Appendix~\ref{ap:cdwvswc}, the correct extension
of the FL theory to the WC case, leads in fact to an {\it increase} of the pining frequency to density
as $n^{\frac3{2}}$. This is due to the fact that
Refs.~\onlinecite{fukuyama_pinning,fukuyama_cdw_magnetic} deals with charge density waves where
the equivalent of  the displacement field $u$ is the phase which is dimensionless. Reinstating the
correct factor $a$ which is the unit of displacement gives an extra factor of the density $n^2$.
The fact that a naive extrapolation of the FL theory is not possible for pinned WC is again clearly
demonstrated by the experiment. The correct behavior (\ref{eq:wpdens})
is $n^{-\frac3{2}}$. So our theory is perfectly compatible with the experimental data.
In addition, other features
such as the non-lorentzian shape of the peaks, are also clearly visible in the data.
In  this experiment, the peak height decreases and peak width increases as the pinning frequency  increases.
This is in qualitative agreement with our predictions for the
pinned WC, as shown in Sec.~\ref{sec:density}.
Compare, for example the qualitative lineshape of Fig.~\ref{fig:bcond} with the data in  Fig.~2 of
Ref.~\onlinecite{li_conductivity_wigner_density}.
Thus the agreement of these experiments with our theory   strongly  supports the
interpretation of the observed phase as a pinned WC.

Furthermore, our results suggests other possible experiments to check for further experimental
signatures of the WC.
The middle frequency range (i.e. $\omega_p \ll \omega \ll \omega_c$) could be compared with the theoretical
formula (\ref{cxx1}). Working at frequencies above the pinning peak has two advantages: (i) the
system is more free from thermal fluctuations, since frequency becomes the dominant energy scale;
(ii) as explained in Sec.~\ref{sec:limitations} the effects of the topological defects are
totally unimportant, so one expects our variational results to be a very good description of
this regime. Another useful probe of the existence of the crystal, is given by difference in behavior of the
conductivity as a function of the both the density and the magnetic field. In the liquid phase,
all quantities are functions of the filling fraction $\nu = n/B$, but do not depend separately
on these two quantities. The striking difference between the $n$ and $B$ dependence seen in
experiment is thus another strong indication of the presence of the crystal. It would thus
be interesting to systematically investigate whether a scaling in $n/B$ is obeyed or not. Such a scaling could
then signal the melting of the crystal.

\section{Comparison with other theories}\label{sec:otherth}

Let us now compare our work with other theoretical work on the conductivity of
the Wigner crystal.

As already mentioned in the previous sections, our work and
 the pioneering work of FL for the pinning of CDW share the same
starting point. However, it differs from the
FL-type analysis in the following essential ways:
(i) the GVM permits  a complete and consistent calculation of the dynamical
conductivity. This allows for a precise calculation of the pinning frequency and
the disorder induced dissipation (see Sec.~\ref{sec:summary}); (ii) the GVM  bypasses the problem
of the unknown fitting parameter that occurs in the Self Consistent Born Approximation
\cite{fukuyama_pinning}, and thus provides an unambiguous determination of the conductivity.
The GVM allows for a proper description of the inherent glassy aspects of the system, and
in particular, of the metastable states;
(iii) more importantly, to treat correctly the case of the pinned WC, it is necessary, contrarily
to the CDW case, to retain variations of disorder at lengthscales smaller than the particle
size. This point was overlooked in more recent applications of the FL theory to the case
of the pinned WC \cite{normand_millis_wigner,li_conductivity_wigner_density}.
As has been extensively discussed in the preceding sections, this leads to radically different dependence of the
pinning frequency on the magnetic field and density.

The question of the linewidth of the pinning peak in conductivity
was also addressed in two series of works.
Refs.~\onlinecite{yi_pinning_wigner,fertig_pinning_wigner} map the
the crystal onto an effective bosonic model. Within a mean field
approach, they obtained exponentially narrow widths of the pinning
peak. Since the computed width is much  narrower
than  observed, extraneous modes of broadening need to be invoked to explain the experimental results. In
Ref.~\onlinecite{fogler_pinning_wigner}, an expression for
$\sigma_{xx}$ was obtained in the limit of pinning frequencies
much smaller than the cyclotron frequency. From the conductivity
expression a formula, linking  the half width of the peaks with
the low frequency part of the conductivity was obtained giving
$\Delta \omega_p /\omega_p \sim (\omega_p/\omega_c)^s$, where $s$
is the exponent governing the low frequency part of the
conductivity. Using a quite different method,
the authors of Ref.~\onlinecite{fogler_pinning_wigner} unknowingly rederived formulas
for the conductivity (their formulas (62) and (68) and the
associated energy scales) that are identical the ones obtained by
us via the GVM in Ref.~\onlinecite{chitra_wigner_hall}. Not
surprisingly, Ref.~\onlinecite{fogler_pinning_wigner} find the
same energy scales. The only difference is the dissipation
function $I(\omega)$ which is replaced in
Ref.~\onlinecite{fogler_pinning_wigner} by an arbitrary
dissipation function (denoted $f$ and taken to be purely
imaginary) scaling for low frequencies as $\omega^{2s}$. In our
case the GVM completely fixes the function $I(\omega)$ (with both
a real and imaginary part) which scales as $\omega^1$, thus fixing
$s=1/2$. In Ref.~\onlinecite{fogler_pinning_wigner}, though the exponent
$s$ could not be computed, arguments were given that the value
should be $s = 3/2$, leading to much more narrow peaks than those
stemming from the GVM. Although, it is in principle possible for
the GVM to miss a part of the excitations (see e.g. the discussion
in Ref.~\onlinecite{giamarchi_columnar_variat,giamarchi_quantum_revue})
we argue that for the pinned WC the exponent $s=1/2$ is correct for the following
reasons: (i) while it is conceivable for the GVM to miss part of
the excitations (such as solitons) this  oversight should lead to an {\it
underestimate} of the low frequency part of the conductivity,
while an exponent $s=3/2$ would imply that the GVM
overestimates the conductivity; (ii) in $d=1$, the exponent
$s=1/2$ given by the GVM, is exactly the one needed to reproduce
the correct behavior of the conductivity which can also be
determined exactly \cite{giamarchi_columnar_variat}.
In addition the parameter $s=3/2$ leads to peak
widths that are much smaller than the ones that are observed
experimentally. Perhaps the difference between our results and the
approach followed in Ref.~\onlinecite{fogler_pinning_wigner} to
estimate $s$, can be traced to the fact that
Ref.~\onlinecite{fogler_pinning_wigner} uses essentially a
classical approximation for the WC. The $I_2(\omega) \propto
\omega$ on the other hand  depends on $\hbar$ so is a
quantum effect and will vanish if all quantum effects are thrown
away.

In addition to these approaches based on an elastic description of the system,
Ref.~\onlinecite{wulf_wigner} incorporated a different approach based on the
microscopic model of
an electron gas confined to its lowest landau level.
Though  they  find that  the pinning frequency
increases with the field, they cannot account for the sharp increase in peak amplitudes  nor do they
obtain the correct dispersion of the collective modes.
The authors suggest  the interesting possibility
that electronic single particle excitations might in some indirect way
be responsible for the experimentally seen behavior.  Further work is
required before unambiguous theoretical conclusions can be drawn.

\section{Conclusions} \label{sec:conclusion}

We have presented in this paper a quantitative theory of the pinned Wigner crystal.
We computed static  and a.c. transport properties. We have obtained the complete
frequency dependence of the imaginary and real parts of the
conductivity tensor. We have shown that the a.c. transport properties
and in particular, the features of the pinning peak,
can be reliably extracted from an elastic theory, without having to take
into account topological defects of the pinned WC. We have studied the evolution of
various  dynamical quantities such as the pinning frequency when the
magnetic field and  density are varied.
We have shown that this problem is controlled by
quite different lengthscales as compared to the standard pinned CDW problem.
More specifically,  depending on the strength of the magnetic field, the pinning
frequency can exhibit quite different behaviors as a function of magnetic field.
Our theory presents a consistent explanation of  most of the observed
experimental features, particularly  the recent experiments
\cite{li_conductivity_wigner_magneticfield,hennigan_optical_wigner,li_conductivity_wigner_density}
and thus provides a strong support for the interpretation of the high field phase observed  in the 2DEG as a pinned Wigner crystal.
Note that the deviations in the Hall coefficient \cite{perruchot_wigner_hall,perruchot_wigner_transverse_force}
observed in the pinned phase, are consistent with the predictions of a transverse critical force
\cite{giamarchi_moving_prl,ledoussal_mglass_long} that should exist for a pinned periodic structure, again
in agreement with the interpretation of this phase as a pinned Wigner crystal.
Conductivity and threshold measurements on the zero-field WC phase seen in
Ga-As samples,  as a function of $B$ and $n$
should provide an ideal testing ground for our theory.
Also a better  understanding   of the disorder present in these samples is required  before a realistic
comparison with theoretical models can be made.

Our work prompts  several questions and issues that still need to be
analyzed. An important issue is of course the influence of the anomalous
(fractional quantum hall) liquid phase. As we have argued this should have little
influence deep in the crystal phase, but certainly changes the melting or
the properties close to melting in a way that  still remains to be understood.
This might be a reason for the strange behavior observed at intermediate magnetic
fields. The way the disorder or the elastic moduli could be affected by the
magnetic field in this region  also demands attention.
But the most important question is the effect of topological defects at finite
temperature. As we have shown, they are not important for the a.c. transport.
But most likely they will be central in understanding the d.c. transport.
It is not known whether the dc resistivity has any anomalous temperature
dependence like those seen in experiments where $\rho_{xx}(T) \propto
\exp{(T_0/T)^\nu}$ where $\nu$ is an exponent which is close to $0.5$ in
Si MOSFets and is found to vary with density in the experiments on
the zero field GaAS samples. Other interesting questions concern the scaling with temperature and
electric field seen in the data. Can such a scaling be accommodated
within the pinned WC approach ?
We leave these  directions for future work.

\begin{acknowledgments}
We would like to acknowledge many interesting discussions with
H. A. Fertig, M.M. Fogler, C.C. Li, F.I.B Williams and J. Yoon.
\end{acknowledgments}

\appendix
\section{Effective disorder} \label{ap:cutdelta}
Here, we derive the disorder term present in (\ref{eq:seff}). Averaging (\ref{eq:ham})
over disorder leads to an inter-replica interaction of the form
\begin{eqnarray}
S_{\rm int} &=& - \int dr dr'\int dr_1 dr_2 \Delta_{r_f}(r-r') \overline{\delta}
(r-r_1) \nonumber \\
& & \overline{\delta}(r'-r_2)
\rho_{l_c=0}(r_1,\tau_1)\rho_{l_c=0}(r_2,\tau_2)
\end{eqnarray}
(the  time integrations  are  implicit)
where
\begin{equation}
\rho = \sum_i \delta(r-R_i-u_i)
\end{equation}
is the density for point like particles.
Using (\ref{eq:fourdens}) (with an unbounded summation over $K$) one easily gets
\begin{eqnarray} \label{eq:ugly}
S_{\rm int} &=& - \rho_{0}^{2} of{}\sum_{K_1,K_2}
\int dr dr' \int dr_1 dr_2 \Delta_{r_f}(r-r') \overline{\delta}(r-r_1) \nonumber \\
& &\overline{\delta}(r'-r_2) e^{i[K_1 (r_1+ u(r_1)) - K_2 (r_2 + u(r_2))
 ]}
\end{eqnarray}
Given the nature of the various functions in (\ref{eq:ugly}) and that  $r_1 - r_2 \ll a$ (typically $r_1-r_2 < \max(r_f,l_c)$)
, in the elastic approximation, one can replace $u(r_1,\tau_1)-u(r_2,\tau_2)$ by
$u(r,\tau_1)-u(r,\tau_2)$
where $r=(r_1+r_2)/2$. Then one easily obtains (\ref{eq:seff}) from (\ref{eq:ugly}) with
\begin{equation}
\Delta_K = \Delta(K) \overline{\delta}(K) \overline{\delta}(-K)
\end{equation}
where $\Delta(K)$ and $\overline{\delta}(K)$ are the respective Fourier transforms of
$\Delta_{r_f}$ and $\overline{\delta}$.

\section{Solution of the variational equations} \label{ap:solvar}

Here, we present the derivation of the saddle point equations
(\ref{iwn1}) and (\ref{sigmasum}) and the detailed structure of the
variational solutions.
Using the parametrization of the replica Green's function given in
(\ref{eq:param}), we  minimize the free energy to obtain the following saddle
point equations. The off diagonal elements of $\sigma$ yield
\begin{equation}\label{eq:sadeq}
\sigma_{\alpha \beta}^{a\neq b}= 2 \int
d\tau \cos(\omega_n \tau) V'_{\alpha \beta}(B^{a\neq b})
\end{equation}
where the displacement correlation function $B$ is given by (\ref{dispcor})

\begin{equation} \label{eq:vsadle}
V(B^{a\neq b})=\sum_K \frac{\Delta_K}{4} e^{-\frac12 \sum_{\alpha\beta}
 K_{\alpha}
K_{\beta} B_{\alpha \beta}^{a\neq b}(0, \tau)}
\end{equation}
and the prime denotes derivatives of $V$ with respect to the argument
$B_{\alpha \beta}^{a\neq b}$.
The diagonal components yield another set of equations involving
the connected part of the Greens functions defined
by $G_{c \alpha \beta}^{-1}= \sum_{b} {G^{ab}_{\alpha \beta}}^{-1 }$.
\begin{eqnarray}\label{gc}
 G_{c11}&=& {{Y+F_{11}}\over
 {{(Y+F_{11}) (X+F_{22})-  (Z + F_{12})(-Z+F_{21})}}} \nonumber \\
G_{c22}&=& {{X+F_{22}}\over
 {{(Y+F_{11}) (X+F_{22})-  (Z + F_{12})(-Z+F_{21})}}}
\\
G_{c12}&=& {{Z+F_{12}}\over
 {{(Y+F_{11}) (X+F_{22})-  (Z + F_{12})(-Z+F_{21})}}}
\nonumber \\
G_{c21}&=& {{-Z+F_{21}}\over
 {{(Y+F_{11}) (X+F_{22})-  (Z + F_{12})(-Z+F_{21})}}} \nonumber
\end{eqnarray}
where $1$ and $2$ denote the  coordinates $x$ and $y$.
\begin{eqnarray}
X &=&( \rho_m\omega_n^2+cq^2 + d {{q_x^2} \over q}) \nonumber \\
Y &=&( \rho_m\omega_n^2+cq^2 +  d{{q_y^2} \over q}) \\
Z &=&( \rho_m\omega_c \omega_n+  d{{q_x q_y} \over q})\nonumber
\end{eqnarray}
\noindent
and
\begin{eqnarray}
F_{\alpha \beta} &=& \frac{2\rho_{0}^{2}}{\beta }
\int d\tau (1- \cos (\omega_n \tau))\sum_K \Delta_K
K_\alpha K_\beta \nonumber \\
& & [e^{-K_{\mu} K_{\nu} B^{aa}_{\mu \nu}} +
\sum_{b, (b\neq a)} e^{ -K_{\mu}K_{\nu}B_{\mu\nu }^{ab}}]
\end{eqnarray}
The above equations form a closed and self-consistent set.
These  have to solved in
the limit  where the number of replicas  $n\to 0$.
Typically, there are two generic solutions:
one is the replica symmetric solution and the other is the
replica symmetry broken solution.  The solution  appropriate  to the
problem studied  depends on
the number of space dimensions, the regime of temperature studied
and the nature of the disorder. For the present problem of a weakly
pinned WC, it has
been shown \cite{giamarchi_columnar_variat} that the relevant solution
is the one with replica symmetry breaking (RSB).
Accordingly, we  parametrize  the  replica Green's functions
in terms of hierarchical matrices such that
\begin{eqnarray}
G^{aa}_{\alpha\beta}(q,\omega_n) &=& {\tilde G}_{\alpha \beta}(q,\omega_n) \\
G^{ab}_{\alpha\beta}(q,\omega_n) &=& G_{\alpha\beta}(q,\omega_n,u)
\end{eqnarray}
where the continuous variable $0<u<1$. The displacement correlations $B$ given
by (\ref{eq:discor}) are similarly parametrised, resulting in two functions
$B_{\alpha \beta}(u)$ and ${\tilde B}_{\alpha \beta}$.

An important simplification occurs
when one considers  quantum problems with quenched disorder or
equivalently classical systems with correlated disorder: since
in each realization of the
random potential, the disorder is time independent,
 all quantities off-diagonal in the replica
indices are always $\tau-$
independent \cite{giamarchi_columnar_variat,giamarchi_quantum_revue}.
Using this property, (\ref{eq:sadeq}) now simplifies to
\begin{equation}
\sigma_{\alpha \beta}(\omega_n, u)\equiv \sigma_{\alpha \beta}(u)={{ 2
\beta}
  }
V^{\prime}_{\alpha \beta} \delta_{n,0}
\end{equation}
Thus all off-diagonal replica coupling, and in particular the
replica symmetry breaking (RSB) is confined entirely to  the
$\omega_n =0$ mode.
For the present problem, an additional simplification accrues  from the
fact that (\ref{eq:vsadle})  involves only the
local (i.e. at $r=0$) $B$ and ${\tilde B}$.
These local quantities are isotropic and thus
$B_{\alpha \beta}(0,u) = \delta_{\alpha \beta} B(u)$ and ${\tilde
B}_{\alpha \beta}(0,\tau) = \delta_{\alpha \beta} {\tilde
B}(\tau)$. This isotropy results in the  enormous simplification
\begin{equation}
\sigma_{\alpha\beta}=\sigma \delta_{\alpha \beta}
\end{equation}
As a result,
\begin{equation}
\label{f12} F_{12}=F_{21} =0
\end{equation}
Therefore, the numerator of $G_{c12}$ in (\ref{gc}) is
unaffected by disorder.

We now define  a new variable $[\sigma]=u
\sigma(u) - \int_0 ^u \sigma(v) dv$.  Based on the generic form of the replica
solutions, we introduce the break-point $u_c$  such that $\sigma(u)=\sigma(u_c)$ for $u \geq u_c$.
If we retain only
the lowest harmonic in the cosine term, the solution has one step
RSB with $\sigma(u)=0$ for $ u < u_c$ and
$\sigma(u)=\sigma(u_c)$ for $u \geq u_c$. In terms of the variable $[\sigma]$,
this RSB ansatz takes the form
\begin{eqnarray} \label{boxsig}
\ [\sigma](u)&=& 0 \qquad  u< u_c \\
\ [\sigma](u)&=& \Sigma = {{2 \beta u_c V^{\prime}(B)}  }  \qquad u\geq u_c \nonumber
\end{eqnarray}
Physically, the region $u <u_c$ corresponds to
the  logarithmic regime, characterized by a logarithmic growth of
displacements  and $u>u_c$  is the Larkin regime
where  $[\sigma](u)= \Sigma$ for $u > u_c$.
This is basically a regime where for  length
scales below the Larkin length $R_c$, the physics of the system
can be approximated by that where independent random forces act
on each electron.
However, for a  multi-cosine model there exists  a third kind of behavior
for intermediate length scales. This regime, called the
 random manifold regime,
is  characterized by
a power law growth of displacements, but with an exponent $\nu$ which
is different from that in the Larkin regime.

 The above discussion  can be summarized in technical terms as follows:
the  solution  $[\sigma]=0$
is valid upto a break-point $u_a$  where the
random manifold regime starts manifesting itself. This is characterized
by a $[\sigma] \propto u^{\mu}$  where $\mu$ is the random manifold
exponent. This power law growth of $[\sigma]$ is valid upto a
second break-point $u_c$ beyond which the Larkin kind of behavior
takes over. For single cosine models like CDW, $u_a =u_c$ and the
random manifold regime shrinks to zero.
The absolute value of $\Sigma$ and hence $R_c$,  depend on
the
 the number of Fourier components of the
disorder that is retained in the problem.   Hence it is important to
retain all the harmonics in the density expansion
so as to get the correct   estimate of $\Sigma$.
Here, for the quantities of our interest, though it is necessary to retain
the higher harmonics in the expression for   $\Sigma$,  we do not require the explicit value of $u_{a }$ for our purposes and   and hence it is sufficient to obtain the value of
$u_{c}$ alone.

We now switch to the transverse and longitudinal components described
in
the main body of the paper. These render the saddle point equations
more
tractable
 and yield the following connected Green's functions
\begin{eqnarray} \label{gctl}
G_{cT}^{-1}&= &(cq^2 + {\rho_m}\omega_n^2) +F +
{{{\rho_m^2}\omega_n^2 \omega_c^2} \over
{(cq^2+dq+ {\rho_m}\omega_n^2 +F)}} \\
G_{cL}^{-1}&= &(cq^2 + dq+{\rho_m}\omega_n^2) +F +
{{{\rho_m^2}\omega_n^2 \omega_c^2} \over {(cq^2 +\rho_m\omega_n^2
+F)}}
\nonumber \\
G_{cLT}^{-1}&= &{\rho_m\omega_n \omega_c} +{{(cq^2
+\rho_m\omega_n^2+F) (cq^2+dq +\rho_m\omega_n^2+F)} \over
{\rho_m\omega_n \omega_c}} \nonumber
\end{eqnarray}

Using (\ref{boxsig}),
the disorder dependent part of (\ref{gctl}) can always be written as
\begin{equation}
  F=I(i\omega_n) + \Sigma (1 -
\delta_{n,0})
\end{equation}
Re-casting all the saddle point equations in terms of the longitudinal and
transverse coordinates,
we find that in the $n\to 0$ limit these reduce to  equations determining $u_c$, $\Sigma$ and
$I(i\omega_n)$ \cite{giamarchi_columnar_variat}. $\Sigma$ is  determined by
\begin{eqnarray} \label{self1}
1&=&-2 V^{\prime \prime}(B(u_c))\int_{\bf q} \left[{1 \over {( cq^2 +
\Sigma)^2}} + {1 \over { (cq^2 + dq + \Sigma)^2}}\right]
\end{eqnarray}
where the off diagonal element $B (u_{c})$ is given by:
\begin{eqnarray}
\label{b_offdiag}
B(u_c)&=& { \over
\beta} \int_{\bf q} \left[ \sum_{n \neq 0} (G_{cL}(q,\omega_n)+
G_{cT}(q,\omega_n) ) \right.\nonumber \\ &+&\left. {1 \over {cq^2 +dq+\Sigma}}
+ {1 \over {cq^2 +\Sigma}}\right]
\end{eqnarray}
and retaining only the fundamental harmonic $K_0$ in $V$, we obtain
\begin{equation} \label{resuc}
\beta u_c = {{  K_0^2} \over {8\pi c}}
\end{equation}
We draw attention to
the fact that $u_c$ is unaffected by the magnetic field and has the
same value as that for  a classical  system with coulomb
interactions and pointlike disorder.
Lastly,
\begin{equation} \label{iw0}
I(i\omega_n)= {2 \over }\int d\tau (1-\cos(\omega_n
\tau)) [V^{\prime}(\tilde B(\tau)) - V^{\prime}(B (u_{c}))]
\end{equation}
\noindent
In (\ref{self1}-\ref{iw0}), the local diagonal correlation
\begin{eqnarray}
\tilde B(\tau) &=&
\frac12 \overline{\langle (u(0,\tau)-u(0,0))^2\rangle }\nonumber
\\
&\equiv&  \frac12
[{\tilde B}_L(\tau)+{\tilde B}_T(\tau)]
\end{eqnarray}
where
\begin{equation} \label{btbl}
{\tilde B}_{T,L}(\tau) = {{2 } \over \beta} \int_q \sum_n
G_{cT,L} [1- \cos(\omega_n \tau)]
\end{equation}

In these expressions,  $G_{cL,T}= (G_{cL,T}^{-1})^{-1}$.
Since $B(u_c)$ depends on $\sum_{n\neq 0} G_c(\omega _n)$ we see from
(\ref{buc}) that $\Sigma$ is a dynamical quantity and hence
 can depend on the cyclotron frequency
$\omega_c$.
To solve these equations, we work with a finite momentum cut off. Fortunately,
for the case of the WC, the bulk modulus being much greater than
the shear modulus,  provides a natural cutoff
$\Lambda=\frac{d}{c}$.

Doing the momentum integral in (\ref{self1}),
the equation for $\Sigma$ can be rewritten as:
\begin{equation}\label{asigmasum}
\Sigma = \sum_{ K} \rho_0^2\Delta_{ K}
{{{ K}^4}\over {4\pi}}   \exp - {1 \over 2} { K}^2 B(u_c)
\end{equation}
 We then take  the semi-classical limit $\hbar \to 0$ in (\ref{iw0}),
This limit allows us to  expand the exponentials in $V$. Using
(\ref{resuc}) and substituting the expressions for $B (u_{c})$ and $\tilde{B}$
\begin{widetext}
\begin{equation}\label{aiwn1}
I(i\omega_n)= -2 V^{\prime\prime}(0) \int_{\bf q} \left[{1 \over
{cq^2 + \Sigma}} +{1 \over { cq^2 +dq+\Sigma}} - {{2(cq^2 +
\omega_n^2 +I(i\omega_n) + \Sigma) + dq} \over {(cq^2 +
\rho_m\omega_n^2+dq + I(i\omega_n) + \Sigma) (cq^2 +
\rho_m\omega_n^2+ I(i\omega_n) + \Sigma)}}\right]
\end{equation}
\end{widetext}
We have thus reduced the entire set of saddle point
equations (\ref{eq:sadeq},\ref{gc}) to two equations for $\Sigma$
and $I (i\omega_{n})$  [(\ref{asigmasum}) and (\ref{aiwn1})] which can then be solved self
consistently to obtain the physical properties of interest.

\section{Fukuyama-Lee results for the WC and CDW} \label{ap:cdwvswc}
In this appendix, we clarify the connection between the results presented
here, and those of Ref.~\onlinecite{fukuyama_pinning} as applied to the WC.
The Fukuyama-Lee results have been misquoted in the literature.  The
 pinning frequency  of a CDW or WC  obtained  in  Ref.\onlinecite{fukuyama_pinning}
is given by
\begin{equation}
\omega_{p0}= { {c R_{a}^{-2}}\over {n\omega_{c}}}
\label{pin}
\end{equation}
\noindent
 The shear modulus $c \propto n^{\frac3{2}}$  and
$R_{a} \propto n^{1/2}$ for CDW  and for the WC,  $R_{a}\propto
n^{-\frac{1}{2}}$. This difference, stems from the fact that for CDW,
the equivalent of  the displacement field $u$ is  a phase variable, which is dimensionless.
Within this approach, the density dependence  is given
$\omega_{p0}\sim n^{3 \over 2}$ for the WC and $\omega_{p0}
\sim n^{-{1 \over 2}}$ for the CDW. This last result was erroneously used to
compare the experimental data with a pinned WC in Ref.~\onlinecite{li_conductivity_wigner_density}.
As our work has emphasized, the correct pinning frequency is determined
by $R_{c}$ and this leads to
\begin{equation}
\omega_{p0} \propto {n}^{-3/2}
\end{equation}
for the WC.

%\bibliography{totphys,wig_th}

\end{document}